\documentclass[useAMS,usenatbib,usegraphicx]{mn2e}
\usepackage{txfonts}
\usepackage{color}
\def\diff{\mathrm d}

\title[Metallicity of SNIa and their host galaxies]{Is the metallicity of their hosts a good
measure of the metallicity of Type Ia supernovae?} 
\author[E. Bravo, C. Badenes]{E. Bravo$^1$\thanks{E-mail: eduardo.bravo@upc.edu} and C.
Badenes$^{2,3}$\thanks{E-mail:
carles@wise.tau.ac.il}\\
$^1$Dept. F\'\i sica i Enginyeria Nuclear, Univ. Polit\`ecnica de
              Catalunya, Carrer Comte d'Urgell 187, 08036 Barcelona, Spain\\   
$^2$Benoziyo Center for Astrophysics, Weizmann Institute of Science, Rehovot 76100, Israel\\
$^3$School of Physics and Astronomy, Tel-Aviv University, Tel-Aviv 69978, Israel}
\begin{document}

\date{Received ; accepted }

\maketitle

\begin{abstract}
The efficient use of Type Ia supernovae (SNIa) for cosmological studies requires knowledge of any
parameter that can affect their luminosity in either systematic or statistical ways. Observational
samples of SNIa commonly use the metallicity of the host galaxy, $Z_\mathrm{host}$, as an estimator
of the supernova progenitor metallicity, $Z_\mathrm{Ia}$, that is one of the primary factors
affecting SNIa magnitude. Here, we present a theoretical study of the relationship between
$Z_\mathrm{Ia}$ and $Z_\mathrm{host}$. We follow the chemical evolution of homogeneous galaxy
models together with the evolution of the supernova rates in order to evaluate the metallicity
distribution function, $\mathrm{MDF}(\Delta Z)$, i.e. the probability that the logarithm of the
metallicity of a SNIa exploding now differs in less than $\Delta Z$ from that of its host. We
analyse several model galaxies aimed to represent from active to passive galaxies, including dwarf
galaxies prone to experience supernova driven outflows. We analyse as well the sensitivity of the
MDF to the most uncertain ingredients of our approach: IMF, star-formation law, stellar lifetime,
stellar yields, and SNIa delay-time distribution (DTD). Our results show a remarkable degree of
agreement between the mean $\bar{Z}_\mathrm{Ia}$ in a galaxy and its $Z_\mathrm{host}$ when they
both are measured as the CNO abundance, especially if the DTD peaks at small time delays, while the
average Fe abundance of host and SNIa may differ up to 0.4-0.6 dex in passive galaxies. The
dispersion of $Z_\mathrm{Ia}$ in active galaxy models is quite small, meaning that
$Z_\mathrm{host}$ is a quite good estimator of the supernova metallicity. Passive galaxies present
a larger dispersion, which is more pronounced in low mass galaxies. 
We present a procedure to generate random SNIa metallicities, given the host metallicity.
We also discuss the use of
different metallicity indicators: Fe vs. O, and gas-phase metallicity vs. stellar
metallicity. Finally, the results of the application of our formalism to a galactic
catalogue (VESPA) suggest that SNIa come, in average, from small metallicity progenitors both at
low redshifts (contrary to expectations) and in galaxies with high star-formation activity. 
In spite of large uncertainties in the metallicities derived from the catalogue, the gross trends
of $\bar{Z}_\mathrm{Ia}$ vs. $Z_\mathrm{host}$ obtained from VESPA for different galaxy types are
roughly consistent with our theoretical estimates.
\end{abstract}


\section{Introduction}

Metallicity is one of the few progenitor attributes that can affect the luminosity of Type Ia
supernovae (SNIa), with important consequences for their use as cosmological standard candles
\citep{dom01,tim03,kas09}. Recently, \citet{bra10} analysed the metallicity as a source of
dispersion in the light-curve width-luminosity relationship and found that deriving
supernova luminosities from light curve shapes without accounting for the supernova metallicity
might
lead to systematic errors of up to 0.5 mag. From an observational point of view, the dependence of
SNIa luminosities on metallicity might originate a systematic error of $\sim9$\% in the
measurement of the dark energy equation of state $w$ \citep{gal08,sul10}, comparable to the
current statistical uncertainties. \citet{sul10} proposed to include the host properties (galaxy
mass and metallicity) into the SNIa light-curve fitters in order to correct for these systematic
errors.

Up to now, attempts to measure the metallicity directly from supernova observations have been
scarce and their results uncertain \citep{len00,tau08}. Measuring the metallicity, $Z$, from the
X-ray emission of supernova remnants is a promising alternative but as yet has been only applied to
a single supernova \citep{bad08a}, although there are prospects to extend this analysis to further
remnants \citep{bad10,yam10}. An alternative venue is to estimate the supernova metallicity as the
mean $Z$ of its environment \citep{bad09} - this can be done for statistically significant samples
of SNe in nearby galaxies by taking advantage of the long evolutionary timescales of their
supernova remnants \citep{bad10b}. 

Excluding these studies based on individual objects or supernova
remnants, the vast majority of our knowledge about the metallicity of Type Ia SN progenitors has
been assembled by measuring the bulk properties of their \textit{entire} host
galaxies. \citet{ham00} looked for galactic age or metal content
correlations with SNIa luminosity in star-forming hosts, but their results were ambiguous.
\citet{gal05} studied the correlations between SNIa properties and galaxy metallicity or,
more precisely, its oxygen abundance as determined from emission lines in the integrated spectra of
star-forming hosts. \cite{ell08} looked for systematic trends of SNIa UV spectra with metallicity
of the host galaxy, and found that the spectral variations were much larger than predicted by
theoretical models. 
\cite{pri08b} studied the metallicity of all the star-forming galaxies in the Sloan Digital Sky
Survey (SDSS) catalogue with redshifts in the range $0.01<z<0.04$ that hosted SNe of some kind and
noted that SNIa occur in a wide range of metallicities, as low as 0.25 solar.
\cite{coo09}, using data from the SDSS and Supernova Survey concluded
that prompt SNIa are more luminous
in metal-poor systems. \citet{gal08} and \citet{how09}, using different methodologies to estimate
the metallicity of SNIa hosts, arrived to opposite conclusions with respect to the dependence
of supernova luminosity on $Z$, although these discrepancies might have been solved recently
\citep{kel10}. 
We note that most of the previous work has focused on comparisons between the empirical
observationally-determined relationship between the {\it metallicity of hosts} and their
SNIa luminosities on the one hand \citep[e.g.][]{how09} and theoretical
predictions of correlations between the {\it supernova metallicity} and luminosity on the other
hand \citep[e.g.][]{tim03}, thus assuming implicitly that both metallicities are somehow tied.

Estimating the metallicity of an entire galaxy is a complex task, involving many different
observational and theoretical challenges. On the observational side, different tracers exist for
both gas-phase and stellar metallicities, each with their own advantages and disadvantages
\citep{bin98}. On the theoretical side, complex modelling of the observations is often required to
arrive at a numerical estimate of the metallicity, which introduces biases and uncertainties that
depend on the specific technique used \citep[e.g.][]{con09}. \citet{gal08} measured the host
metallicity from absorption lines of Fe at
two wavelengths, thus testing the gas-phase in early-type galaxies, using single-age
single-metallicity stellar population models. Then, to derive metallicity, they assumed solar
elemental ratios in the galactic interstellar medium (ISM). \citet{how09}
determined the host metallicity indirectly. In a first step, they determined the host mass from
spectral energy distribution fits to galaxy photometry. Next, they applied the well-known
mass-metallicity relationship \citep{tre04}, also valid for the gas-phase. The \citet{tre04}
mass-metallicity relationship was determined in turn through measures of the nebular emission
lines of oxygen in star-forming galaxies. \citet{cid05} compared the stellar metallicities to
those of the gas-phase (measured through the oxygen abundance) for several thousand galaxies in the
SDSS. They found that the relationship between both metallicities is
non-linear, with a shallower than linear dependence of the gas-phase oxygen abundance on the
stellar metallicity, and a large scatter of data points.

The question we want to address in this work is: What is the relationship between the metallicity
of a SNIa and that of its host galaxy? Given the complexity of the problem we
want to address it from a statistical point of view, aiming to determine a distribution function
for the difference of metallicities between the supernova and its host. Our approach is based on
calculations of galactic chemical evolution together with the evolution of the supernova rate.
In order to understand which the best estimators of SNIa metallicity are we study the correlations
between SNIa metallicity and different observational tracers of host metallicity: Fe vs. O in the
gas-phase and stellar vs. gas-phase metallicity. In order to keep simplicity, we use homogeneous
one-zone galactic chemical evolution models, disregarding spatial variations within each galaxy.
Taking into account the spatial variations of metallicity would increase the scatter and
corrections modelled in this paper.

The plan of the paper is as follows. In Section~\ref{models}, we describe the theoretical framework
we adopt in order to derive the statistical properties of the metallicities of SNIa compared to
that of their host galaxies. In the next section, we analyse the results, with particular emphasis
on the reliability of different host metallicity measures as estimators of SNIa metallicity. Among
the tracers that we consider, we
rank highest the measure of SNIa metallicity through the gas-phase metallicity of its host. We pay
special attention to the influence of the assumed Delay Time Distribution (DTD) of SNIa. It turns
out that
the metallicity of an average SNIa is closer to that of its host if the DTD peaks at small
time delays, as the DTD of \citet{mao10b}, than if the SNIa rate follows the rate of formation
of white dwarfs, as proposed by \citet{pri08}. In
Section~\ref{catalogs} we explore the possibility of using current galactic catalogues, such as
SDSS and VESPA, to derive the metallicity distribution function of SNIa. Finally, in
Section~\ref{toconclude} we draw our conclusions. 

\section{The chemical evolution of galaxies and their SNIa}\label{models}

\subsection{SNIa Metallicity Distribution Function}

We define the metallicity probability density of SNIa, $P'(\Delta Z)$ as the probability that a
SNIa comes from a progenitor whose metallicity logarithm differs from the
logarithm of its host metallicity, $Z_\mathrm{host}$, in:

\begin{equation}
 \Delta Z \equiv\log\left(Z_\mathrm{Ia}/Z_\mathrm{host}\right)\,.
\label{eqdz}
\end{equation}

\noindent Similarly, we define the metallicity distribution function (MDF) of
SNIa as the cumulative metallicity probability, i.e. the probability
that the difference between the logarithm of the metallicity of the SNIa progenitor and that of the
host is smaller than
$\Delta Z$ (note that this means $\Delta Z$ larger in absolute value in the usual case that the
SNIa metallicity is smaller than the host's one),

\begin{equation}
 \mathrm{MDF}\left(\Delta Z\right) = \int_{-\infty}^{\Delta Z} P'\left(\Delta Z\right)
\mathrm{d}\Delta Z\,.
\end{equation}

\noindent We give additional details about the procedure we use to compute $P'\left(\Delta
Z\right)$ at the end of Section~\ref{sectinf}. We further define the ``metallicity
correction'', $\Delta\bar{Z}$, as the mean value of $\Delta Z$, i.e.:

\begin{equation}
 \Delta \bar{Z} \equiv\overline{\log}\left(Z_\mathrm{Ia}/Z_\mathrm{host}\right)\,.
\end{equation}

The present rate of SNIa explosions in a given host, $R_\mathrm{Ia}(t)$, is the convolution over
time of the Star Formation Rate (SFR), $S$, and the DTD, $\Psi$: 

\begin{equation}
R_\mathrm{Ia} = \int_0^t S(t') \Psi(t-t') dt' \label{eqria}\,.
\end{equation}

\noindent The probability that a SNIa that explodes now (at time $t$) comes from a progenitor born
at time $t' = t - \tau$ is:

\begin{equation}
 P_\mathrm{Ia} \left(\tau\right) = S(t-\tau)\Psi(\tau)/R_\mathrm{Ia}(t)\,, \label{eqpia}
\end{equation}

\noindent where $\tau$ is the time spanned since progenitor birth to the moment the host is being
observed, which we will call look-back time, and $t$ is the cosmic time. Finally, the knowledge of
$P_\mathrm{Ia}$ vs. time and the metallicity evolution, $Z(t)$, allows to construct the MDF of
SNIa. 

We have used two different prescriptions for the DTD: those of \citet{pri08} and \citet[][in the
following M10b]{mao10b}. As we will show later in Section~\ref{sectdtd}, the DTD is the most
influential ingredient for the determination of the MDF of SNIa.  

\citet[][in the following P08]{pri08} found that the observational rate of SNIa is a constant
fraction, $\sim1\%$ of the stellar death rate and independent of the SFR. Later, \citet{ras09} used
a different method to confirm the results of P08, i.e. the rate of SNIa traces the rate
of formation of white dwarfs (WD), with a constant efficiency factor:

\begin{equation}
R_\mathrm{Ia} \propto 
 R_{WD} = \int_{t(M_\mathrm{max})}^t S(t') \phi(M_\tau) \left(\frac{dM}{dt}\right)_\tau dt'\,,
\label{eqrwd}
\end{equation}

\noindent where $\phi = \left(dN/dM\right)$ is the Initial Mass Function (IMF), 
$\left(\frac{dM}{dt}\right)_\tau$ is
the inverse of the derivative of the lifetime of a star of mass $M_\tau$ whose lifetime is
precisely $\tau$, and the integral extends down to the evolution time of the maximum mass of a star
that ends as a WD, $M_\mathrm{max}$. In this formulation, the probability that a SNIa has an age
(i.e., its progenitor's age) $\tau$ is:

\begin{equation}
 P_\mathrm{Ia} \left(\tau\right) \propto S(t-\tau) \phi(M_\tau) \left(\frac{dM}{dt}\right)_\tau \,.
\label{eqpia2}
\end{equation}

On the other hand, when the DTD of M10b is applied the probability that a SNIa
has an age $\tau$ is obtained right from Eq.~\ref{eqpia}:

\begin{equation}
 P_\mathrm{Ia} \left(\tau\right) \propto S\left(t-\tau\right) \Psi(\tau)\,.
\label{eqpia3}
\end{equation}

\subsection{Basic ingredients of the chemical evolution model}

We adopt a Salpeter IMF \citep{sal55} with a small-mass cut-off of
$M_\mathrm{bot}=0.07$~M$_{\sun}$ and a
high-mass cut-off of $M_\mathrm{up}=50$~M$_{\sun}$, and the stellar lifetime law proposed by
\citet{tal71}: 
\begin{equation}
 t(M) = 12 M^{-2.5} + 0.005\,,\label{eqtau}
\end{equation}
\noindent with mass in $M_{\sun}$ and time in
Gyr. In Section~\ref{sectsens} we explore the sensitivity of our results to other choices for the
IMF \citep{cha03,kro07} and the stellar lifetime law \citep{buz02}. 

One of the main ingredients necessary to determine the MDF is the prescription for the stellar
yields, including supernova processing. Here, we adopt the yields provided by \citet{kob06} for
stars from 13 to 40~M$_{\sun}$ and metallicities from 0 to 0.02. Stars in the above mass range
are the main production site of the CNO elements. The \citet{kob06} nucleosynthesis calculations
take into account the entire lifetime of the stars, from zero-age main sequence to the supernova or
hypernova explosion, and provide the mass ejected in the explosion. We assume that the mass lost
by the star during its hydrostatic evolution returns to the ISM the same metal fraction with which
the star was born. 

The yield tables of \citet{kob06} provide the ejected mass of each element, $m_j$, as a function
of the star mass, $M_i$, and metallicity, $Z_i$: $m_j=m_j\left(M_i,Z_i\right)$. In most of our
calculations we interpolate the yields as a function of metallicity, according to the metallicity
of the ISM at the time of stellar birth. As in \citet{nom06}, we adopt a fraction of hypernovas
$\varepsilon_\mathrm{hn}=0.5$. 

In order to simplify, we apply the instantaneous recycling approximation to massive stars. We
integrate the yields between a minimum mass of 10~M$_{\sun}$ and a maximum mass of
$M_\mathrm{up}=50$~M$_{\sun}$, thus obtaining a mean yield, $\bar{y}_j$. In this work, we use two
measures of the metallicity, $Z$, either the 
CNO elements, whose mean yield is $\bar{y}_\mathrm{CNO}$, or all the elements with atomic
number larger than five (we will refer to this last group generically as 'Fe' in the following,
and they will be our default choice for the measure of metallicity, $Z$, unless otherwise stated),
whose mean yield we generically denote as $\bar{y}_\mathrm{Fe}$\footnote{For SNIa theory both
measures of metallicity are meaningful. On one side, the total neutron excess in the exploding
white dwarf, related to the overall metallicity of the progenitor, plays a role in the timescale of
electron captures that control the fraction of incinerated mass that goes to radioactive $^{56}$Ni
and empowers the light curve. On the other side, the CNO elements present in the progenitor turn
into $^{22}$Ne in the white dwarf, which has a relevant role as a catalyst of the thermonuclear
reactions in SNIa and may also influence the brightness of the supernova \citep[for a
discussion see][and references therein]{bra10}.}. The
integration is performed numerically by dividing the full mass interval in bins centred in the
mass of each of the models present in the tables given by \citet{kob06}, then weighting the
respective yield by the IMF of mass $M_i$ and multiplying by the width of the mass interval. 

With respect to stars of mass below 10~M$_{\sun}$, we assume that the difference of mass between
the zero-age main sequence star and its end-of-life compact remnant is returned to the ISM with
the same fractional composition of metals it had at birth. Both, the fraction of the mass of a star
that is returned to the ISM during its hydrostatic evolution, $E_\mathrm{h}(M)$, and the total
fraction of the mass of a star that is returned to the ISM (excluding SNIa), $E(M)$, have been
taken from \citet{cat09} for stars of $M<10$~M$_{\sun}$, and from \citet{kob06} for stars of larger
masses. Stars ending as SNIa return an additional mass to the ISM equal to the
Chandrasekhar mass and composed 100\% of metals, with negligible amounts of CNO elements.

\subsection{Galactic models with mass inflow}\label{sectinf}

To study the statistical properties of the MDF of SNIa we include into our chemical evolution
models both a time-dependent SFR and an infall law that
account for the interaction with the galactic halo. For the SFR we adopt the Schmidt law as
implemented by \citet{kob06}:

\begin{equation}
 S(t)=m_\mathrm{g}(t)/\tau_\mathrm{sfr}\,,\label{eqsfr}
\end{equation}

\noindent where $\tau_\mathrm{sfr}$ is the timescale of star formation, and $m_\mathrm{g}(t)$ is
the mass of gas in the ISM at time $t$. In the literature, there are other
prescriptions for the SFR law. For instance, \citet[][see also \cite{gav02,gre09}]{san86} proposed
to use a universal exponential SFR as a function of time:

\begin{equation}
 S(t)\propto \frac{t}{\tau_\mathrm{sfr}^2}\exp\left[-\left(t/\tau_\mathrm{sfr}\right)^2\right]\,,
\end{equation}

\noindent with $\tau_\mathrm{sfr}\sim2-4$~Gyr for early-type galaxies,
$\tau_\mathrm{sfr}\sim10-13$~ Gyr for spirals, and $\tau_\mathrm{sfr}\sim25$~Gyr for dwarf
irregulars. However, the use of the Sandage SFR law can lead to non-positive gas mass if the
constant of proportionality is not accurately chosen and/or other ingredients such as
inflow/outflow are present in the model. It is for this reason that we prefer to adhere to the
Schmidt law of SFR.

To include the infall of material from the galactic halo we follow the conservative
model of \citet{kob00} and assume that the total mass of the system formed by gas, stars
and halo is constant and equal to $M_{\mathrm{h}0}$. Initially, the whole mass is in the halo,
and the mass infall rate as a function of time is: 

\begin{equation}
 F(t) = \frac{M_{\mathrm{h}0}}{\tau_\mathrm{inf}}\exp\left[-t/\tau_\mathrm{inf}\right]\,,
\end{equation}

\noindent where $\tau_\mathrm{inf}$ is the timescale of infall. \citet{kob00} propose values
of $\tau_\mathrm{inf}=0.1$~Gyr as representative of elliptical galaxies, 10.9~Gyr to account for
Galaxies of type Sbc/Sc, and 60.5~Gyr to represent types Scd/Sd. In this work, we use
$\tau_\mathrm{sfr}$ and $\tau_\mathrm{inf}$ as free parameters in order to obtain models for
different types of galaxies

The set of differential equations to integrate in these models is the following one (in addition
to Eqs.~\ref{eqpia2} or \ref{eqpia3}):

\begin{eqnarray}
 \dot{m}_\mathrm{g} & = & - S(t) + F(t)  +
R_\mathrm{Ia}M_\mathrm{Ch} \nonumber \\
& & + \int_{M(t)}^{M_\mathrm{up}}{\phi(M)S(t-\tau_M)E(M)M \mathrm{d}M}\,, 
\label{eqmg}
\end{eqnarray}

\begin{eqnarray}
 \dot{m}_j & = & - S(t)\left[X_j(t) - \bar{y}_j\right]  +
R_\mathrm{Ia}M_\mathrm{Ch} \nonumber \\
& & + \int_{M(t)}^{M_\mathrm{up}}{\phi(M)S(t-\tau_M)E_\mathrm{h}(M)X_j(t-\tau_M)M \mathrm{d}M}\,,
\label{eqmj}
\end{eqnarray}

\noindent where $M(t)$ is the inverse of the stellar lifetime function (Eq.~\ref{eqtau}), 
$X_j=m_j/m_\mathrm{g}$ is the mass fraction of elements $j$
in the gas at time $t$, $\tau_M$ is given by Eq.~\ref{eqtau}, $S$ a function of time
through
$m_\mathrm{g}(t)$, the subscript $j$ in Eq.~\ref{eqmj} stands for either 'CNO' or 'Fe', and
$R_\mathrm{Ia}$ is computed using Eq.~\ref{eqrwd} (for the P08 DTD) or Eq.~\ref{eqria} (for the
M10b DTD). In this
equation, we assume that the infall gas is of primordial composition and does not
contribute to the mass of metals. Seemingly, the term $R_\mathrm{Ia}M_\mathrm{Ch}$ is only added
to the equation of evolution of $m_\mathrm{Fe}$, but not to that of $m_\mathrm{CNO}$. The last
term in Eq.~\ref{eqmj} accounts for the fraction of metals in the protostellar nebula that 
returns to the ISM during the hydrostatic stellar evolution. 

If the mass fraction $X_j$ were a monotonic function of time, the metallicity probability density
might be obtained directly by combining Eqs.~\ref{eqpia}, \ref{eqmg}, and
\ref{eqmj},\footnote{A normalization constant is missing in Eq.~\ref{eqdirect}. We
have omitted it to improve readability.}

\begin{equation}
 P'\left(\Delta Z\right) = \frac{\mathrm{d}R_\mathrm{Ia}}{\mathrm{d}\Delta Z} = 
 X_j \frac{P_\mathrm{Ia}}{\dot{X}_j}\,,
\label{eqdirect}
\end{equation}

\noindent where we have used the definition of $\Delta Z$ (Eq.~\ref{eqdz}) and the relationship
between $P_\mathrm{Ia}$ and $R_\mathrm{Ia}$, $P_\mathrm{Ia}=\mathrm{d}R_\mathrm{Ia}/\mathrm{d}t$, 
$X_j$ is the measure of $Z$, and

\begin{equation}
 \dot{X}_j = \frac{\dot{m}_j}{m_\mathrm{g}} - \frac{m_j}{m_\mathrm{g}^2}\dot{m}_\mathrm{g}\,. 
\end{equation}

\noindent However, it is not guaranteed that the metallicity is a monotonic function of time, so
we have to resort to a different procedure to obtain the metallicity probability density. In this
work we have computed $P'\left(\Delta Z\right)$ through a Monte Carlo calculation. First, we solve
Eqs.~\ref{eqpia}, \ref{eqmg}, and \ref{eqmj} to obtain $P_\mathrm{Ia}$ and $X_j$ as functions of
time. Second, we generate randomly the distribution of birth times of SNIa according to
$P_\mathrm{Ia}$, by drawing $N$ random numbers, $\chi_i$, from a uniform distribution and assigning
them birth times $t_i$ such that 

\begin{equation}
\chi_i = \int_0^{t_i} P_\mathrm{Ia} dt\,.
\end{equation}

\noindent For each $t_i$ thus obtained, the corresponding $\Delta Z$ is identified and the
probability density is easily calculated. For this last calculation we have used 100 bins in
$\Delta Z$. The results shown in this paper were generated with $N=100000$ random
numbers per each $j$ each time the chemical evolution of a galaxy was computed. 

\subsection{Models of passive galaxies with mass outflow}

Elliptical galaxies span large luminosity and mass ranges, their light emission is dominated by
red giant stars, and contain no gas \citep[for a recent review see][and references therein]{mat08}.
Ellipticals are metal rich, although their mean stellar metallicity is in the range
$[\mathrm{Fe}]\simeq$ from -0.8 to +0.3 \citep{kob99}, in general the metallicity grows with the
mass of the galaxy. The history of star formation in elliptical galaxies is still controversial,
the two main scenarios being the single starburst model (monolithic scenario) and the hierarchical
model. In the former, elliptical galaxies are assumed to have formed at high redshift through a
short and intense starburst as a result of dissipative collapse of protogalactic gas clouds
followed by a supernova driven galactic wind \citep[e.g.][]{lar74,ari87,mat87,pip04}. In the
hierarchical model, elliptical galaxies form at relatively recent epochs through mergers of gaseous
galaxies, with continuous star formation through a wide redshift range
\citep[e.g.][]{whi78,bau98,kau98,ste02}. In every case, elliptical galaxies should be
characterized by short and intense starburst(s) during which a large number of massive stars
form and explode in a short time interval, releasing a large energy into the gas leading to
massive outflows that disperse the supernova yields into the intergalactic medium, especially for
small mass galaxies.

The question of which model best reproduces the
observational constraints is not settled yet. Since we do not pretend to address all the
complexities of the formation of elliptical galaxies \citep[see, e.g.,][]{pip06}, we have just
chosen to adopt a simple prescription intended to reproduce the chemical evolution in the
monolithic scenario: the so-called
``leaky-box model'' of \citet{har76}. In this prescription, the effect of the supernova
driven outflow is encapsulated in the use of an {\it effective yield}, $y_{i,\mathrm{eff}} =
y_i/(1+c_\mathrm{out})$ \citep{bin98}, where $y_i$ is the usual yield of element $i$,
$y_{i,\mathrm{eff}}$ is a reduced yield that goes to the ISM and  
accounts for all the material lost to outflow motions, and $c_\mathrm{out}>0$ is a
constant. The leaky-box model allows us to simulate elliptical galaxies of different masses, where
large values of $c_\mathrm{out}$ represent low-mass low-metallicity early-type galaxy models. We
have computed models of early-type galaxies with infall timescale and SFR
timescale in the range 0.1-0.5 Gyr \citep[see][for an study of the correlation between both
timescales and the size and luminosity of the galaxy]{pip04}, and have explored a range of
$c_\mathrm{out}$ from 0 (no outflow, a model for giant elliptical galaxies) to 10 (meant to
represent dwarf ellipticals).

\subsection{Sensitivity to the IMF, stellar lifetime, and stellar yields}\label{sectsens}

Our objective is to understand the relationship between the metallicities of SNIa and
those of their hosts. The sensitivity of this relationship to uncertain ingredients of the
theoretical model depends on the way these ingredients affect the host chemistry in comparison to
the SNIa rate. Here we use a simplified treatment of the galactic chemical evolution that allows
obtaining MDF independent of the SFR, and the DTD of P08.

We start by considering the simultaneous evolution of metallicity and SNIa rates in a closed-box
model. We further simplify the model by taking a constant SFR, $S_0$, through the complete
galactic life, and assuming the instantaneous recycling approximation for stars more massive than
10~M$_{\sun}$. 
Within these assumptions, the evolution of the mass of gas and of
the mass of CNO elements in the gas phase, $m_\mathrm{CNO}$, are given by the following
differential equations: 

\begin{equation}
 \dot{m}_\mathrm{g} = - S_0\left[1 - \int_{M(t)}^{M_\mathrm{up}}{\phi(M)E(M)M
\mathrm{d}M}\right]\,,
 \label{eqmgsimple}
\end{equation}

\begin{equation}
 \dot{m}_\mathrm{CNO} = - S_0\left[X_\mathrm{CNO}(t) - \bar{y}_\mathrm{CNO}\right]\,,
 \label{eqmcnosimple}
\end{equation}

\noindent where $\bar{y}_\mathrm{CNO}$ incorporates now the last term in Eq.~\ref{eqmj}. Within the
same assumptions, Eqs.~\ref{eqpia2} and \ref{eqpia3} simplify to:

\begin{equation}
 P_\mathrm{Ia} \left(\tau\right) \propto S_0 \phi(M_\tau) \left(\frac{dM}{dt}\right)_\tau = S_0
\Psi(\tau) \,,
\end{equation}

\noindent and the MDF can be obtained formally from:

\begin{equation}
 \frac{\diff R_\mathrm{Ia}}{\diff m_\mathrm{CNO}} \propto \frac{\Psi(\tau)}{X_\mathrm{CNO}(t) -
\bar{y}_\mathrm{CNO}}\,,
\end{equation}

\noindent using the results of the integration of Eqs.~\ref{eqmgsimple} and \ref{eqmcnosimple}.
The last expression shows that the MDF thus obtained is independent of $S_0$.

The IMF we use is that of \citet{sal55}. The results we obtain are insensitive to the
precise values of the cut-off within the range we have explored: $M_\mathrm{bot} = 0.07 -
0.1$~M$_{\sun}$ and $M_\mathrm{up} = 50 - 80$~M$_{\sun}$.

\begin{figure}
\centering
   \includegraphics[width=9 cm]{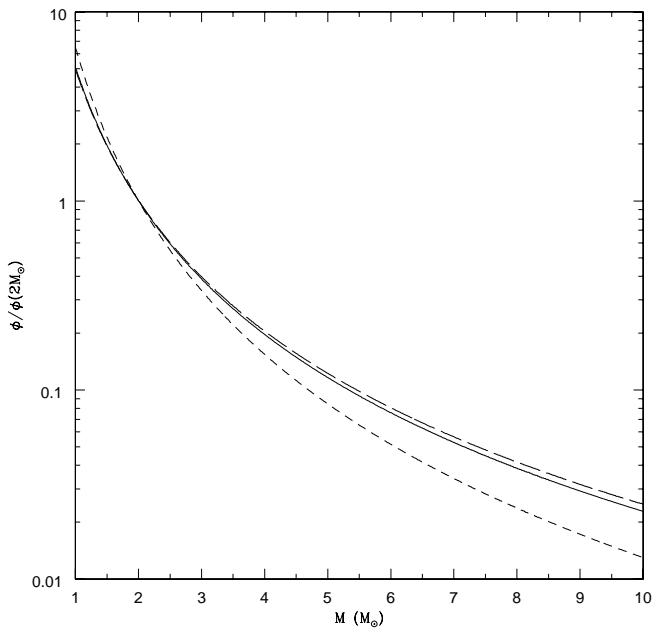}
\caption{Comparison of the IMFs used in the present work in a range of masses of SNIa
progenitors: \citet{sal55} (solid), \citet{kro07} (short dash), and \citet{cha03} (long dash). The
vertical axis gives each IMF as a function of star mass normalized to the same IMF law
for $M=2$~M$_{\sun}$. 
}\label{figimf}
\end{figure}

Besides the IMF of \citet{sal55}, we have considered as well those of \citet{kro07} and
\citet{cha03}. Figure~\ref{figimf} shows the behaviour of these IMF within the range of masses of
SNIa progenitors. In this mass range, the normalized IMF of \citet{sal55} and \citet{cha03} are
nearly indistinguishable, while the IMF of \citet{kro07} gives slightly less weight to the more
massive SNIa progenitors, consequently favouring slightly longer SNIa delay times. These small
differences translate in small variations in the MDF of SNIa. The difference in the metallicity
correction, $\Delta\bar{Z}$, between all these IMF is smaller than $\sim0.08$~dex for the whole
sample of galactic models we have generated.

To address the dependence of our results on the stellar lifetime law, we have repeated the
calculations using the prescriptions of \citet{buz02} instead of \citet{tal71}. In general, the
lifetimes obtained following \citet{buz02} are slightly shorter ($\sim50$\%) than those given by
\citet{tal71}. Although this leads SNIa to follow tighter the SFR (as we are here working
with the DTD of P08), the metallicity correction does
not change by more than $\sim0.07$~dex between both lifetime laws.

In order to test the sensitivity of our results to the prescription of the stellar yields we have
repeated the calculations with the yields $m_j$ given by \citet{kob06} fixed at two
extreme metallicities: either $Z_i=0$ or $Z_i=0.02$. The result is that the MDF of SNIa is
insensitive to the choice of stellar yields. This insensitivity is a consequence of our definition
of the MDF as a function of the ratio of $Z_\mathrm{Ia}$ to $Z_\mathrm{host}$. Hence, even if the
absolute values of both metallicites change with the stellar yield prescription, their ratio is
extremely insensitive to them.

\section{Theoretical results}

In this section, we will present the results obtained with the models including inflow and the
models of passive galaxies. Unless otherwise stated, we solve Eqs.~\ref{eqmg} and \ref{eqmj} from
$t=0$ to $t=t_\mathrm{gal}=13$~Gyr. 

\subsection{Prototype galaxies}\label{proto}

\begin{figure}
\centering
   \includegraphics[width=9 cm]{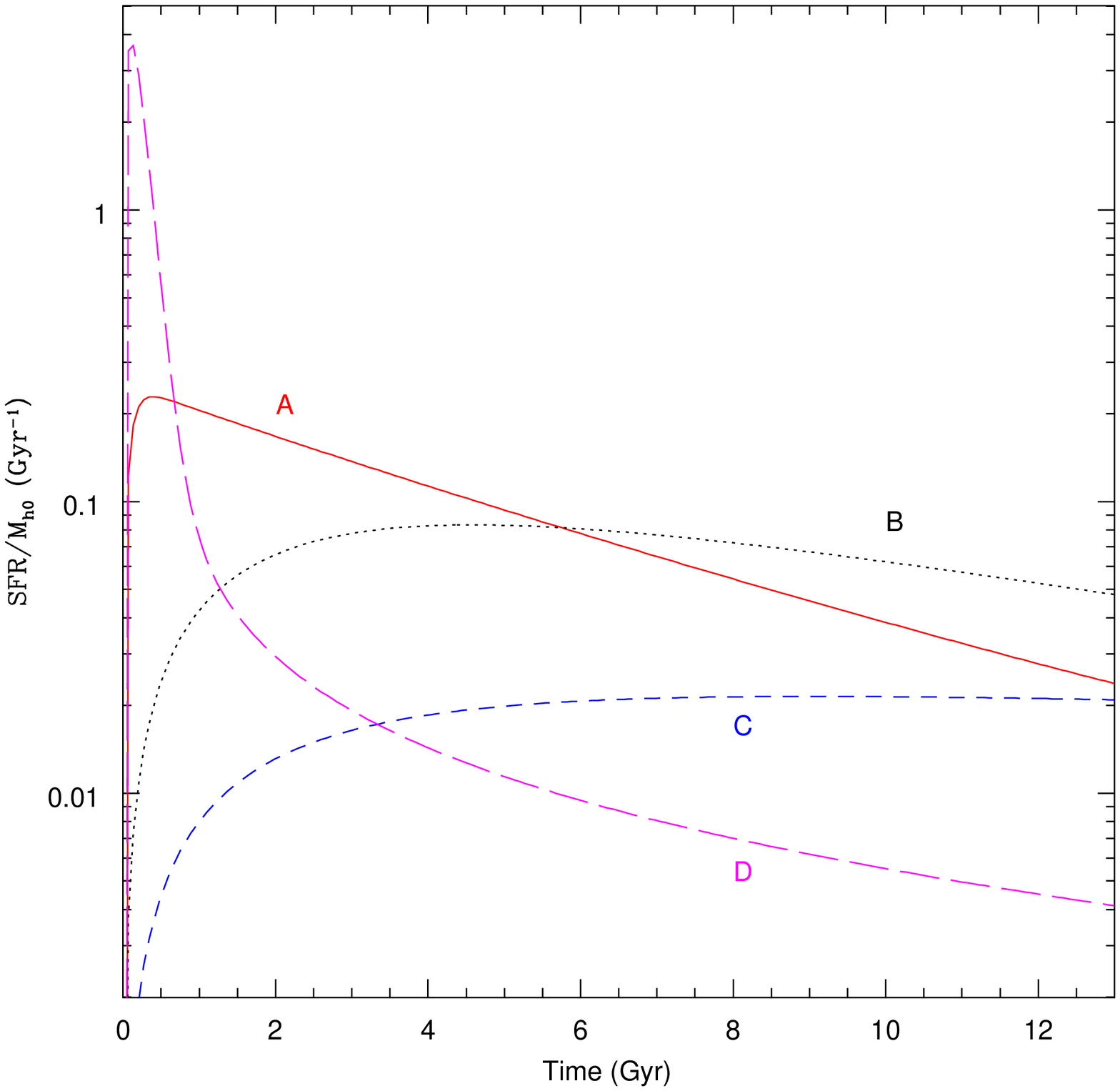}
\caption{
Star formation rates as a function of time for four galaxy models in Table~\ref{tab1}. The
different
curves are meant to represent typical galaxies: model D in Table~\ref{tab1} (long-dashed magenta)
belongs to an early-type galaxy, while models B (dotted black) and C (short-dashed blue) are
representative of late-type galaxies, and model A (solid red) is a transition galaxy
characterized by a short infall timescale and a moderately long star formation rate timescale. The
vertical axis gives the SFR {\it
normalized} by the total mass of the system, $M_\mathrm{h0}$ ($M_\mathrm{h0}$ is also the mass of
the halo at $t=0$ and should be much smaller for irregulars and dwarf ellipticals than
for normal-size galaxies). 
}\label{figsfr}
\end{figure}

\begin{table*}
\begin{minipage}{126mm}
 \caption{Parameters of some typical galaxy models.}
 \label{tab1}
\renewcommand{\thefootnote}{\alph{footnote}}
 \begin{tabular}{@{}lcccccccc@{}}
 \hline
 & \multicolumn{4}{c}{Pritchet-DTD models} & \multicolumn{4}{c}{Maoz-DTD models} \\
 & A & B & C & D & A' & B' & C' & D' \\
 \hline
$\tau_\mathrm{inf}$ (Gyr)$^a$ & 0.1 & 9.0 & 50.0 & 0.1 & 0.1 & 9.0 & 50.0 & 0.1 \\
$\tau_\mathrm{sfr}$ (Gyr)$^b$ & 4.0 & 2.0 & 2.0 & 0.1 & 4.0 & 2.0 & 2.0 & 0.1 \\
DTD\footnotemark & P08 & P08 & P08 & P08 & M10b & M10b & M10b & M10b \\ 
$f_\mathrm{gas}$\footnotemark & 0.09 & 0.09 & 0.04 & 0 & 0.09 & 0.09 & 0.04 & 0 \\
$f_\mathrm{star}$\footnotemark & 0.91 & 0.67 & 0.19 & 1 & 0.91 & 0.67 & 0.19 & 1 \\
$Z_\mathrm{host}$\footnotemark & 0.029 & 0.015 & 0.012 & 0.036 & 0.041 & 0.022 & 0.017 & 0.035 \\
$\Delta\bar{Z}$ (dex)\footnotemark & -0.53 & -0.15 & -0.089 & -0.54 & -0.21 & -0.047 & -0.029 &
-0.33 \\
$\sigma\left(\Delta\bar{Z}\right)$\footnotemark & 0.49 & 0.19 & 0.14 & 0.35 & 0.39 &
0.11 & 0.071 & 0.41 \\
$X\left(\mathrm{CNO}\right)_\mathrm{host}$\footnotemark & 0.017 & 0.0091 & 0.0076 & 0.0085 & 0.017
& 0.0091 & 0.0076 & 0.0085 \\
$\Delta\bar{X}\left(\mathrm{CNO}\right)$ (dex)\footnotemark & -0.48 & -0.12 & -0.070 & -0.13 &
-0.19 & -0.024 & -0.017 & -0.093 \\
$\sigma\{\Delta\bar{X}\left(\mathrm{CNO}\right)\}$\footnotemark & 0.47 & 0.18 & 0.13
& 0.31 & 0.36 & 0.10 & 0.062 & 0.28 \\
 \hline
 \end{tabular}
\footnotetext[1]{Timescale of infall}
\footnotetext[2]{Timescale of star formation}
\footnotetext[3]{P08: \citet{pri08}; M10b: \citet{mao10b}.}
\footnotetext[4]{Fraction of gas mass in the ISM at $t=13$~Gyr with respect to the initial
baryonic halo mass, $M_{h0}$.}
\footnotetext[5]{Fraction of stellar mass at $t=13$~Gyr with respect to the
initial baryonic halo mass, $M_{h0}$.}
\footnotetext[6]{Gas metallicity (mass fraction of all metals) at $t=13$~Gyr.}
\footnotetext[7]{$\Delta\bar{Z}\equiv\overline{\log}\left(Z_\mathrm{Ia}/Z_\mathrm{host}\right)$.}
\footnotetext[8]{Standard deviation of the distribution of SNIa metallicities, MDF.}
\footnotetext[9]{Mass fraction of CNO elements in the gas at $t=13$~Gyr.}
\footnotetext[10]{$\Delta\bar{X}\left(\mathrm{CNO}\right)\equiv\overline{\log}\left[X\left(\mathrm{CNO}
\right)_\mathrm{Ia} /X\left(\mathrm{CNO}\right)_\mathrm{host}\right]$.}
\footnotetext[11]{Standard deviation of the distribution of CNO mass fractions in SNIa.}
\setcounter{footnote}{1}
\end{minipage}
\end{table*}

We start by analysing the behaviour of typical galaxies meant to represent from ellipticals to
late-type galaxies. Table~\ref{tab1} gives their main parameters together with the final
mass fractions of gas and stars with respect to the total mass of the system, the final
metallicity (both as given by all metals and by CNO elements) of the host, and the statistical
properties of the corresponding MDF of SNIa: metallicity correction and standard deviation of the
distribution, $\sigma$.
We computed all the model galaxies discussed in this subsection with no outflow.
Models A to C
are in good agreement with the models in Table 2 of \citet{kob00} for galaxies of types E, Sc, and
Sd (the remaining mass of gas and the mass of stars at $t=t_\mathrm{gal}$ of the models, as well as
the final metallicities agree reasonably well given that the model of
evolution of the rate of SNIa we use are not the same as their).
The SFR of model D is more bursty (Fig.~\ref{figsfr}), and we include it here as a representative
of the early-type galaxy models used in the rest of the paper (even though it has
$c_\mathrm{out}=0$, which would correspond to a massive early-type galaxy). In this
and the next subsection, we present the results obtained with the DTD of P08, while we delay to
Section~\ref{sectdtd} the discussion of the differences with respect to the DTD of M10b.

\begin{figure*}
\centering
   \includegraphics[width=8.8 cm]{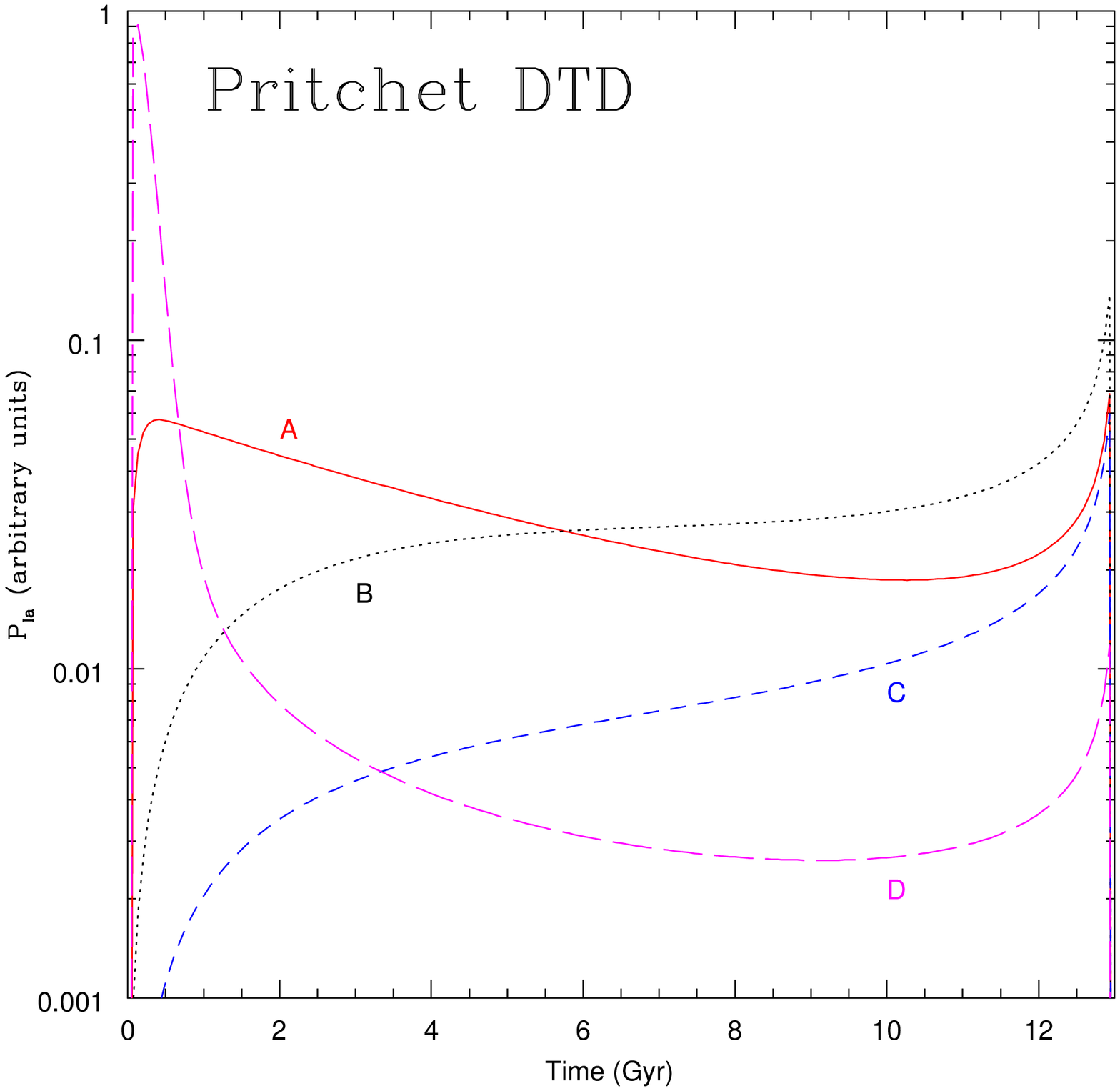}
   \includegraphics[width=8.8 cm]{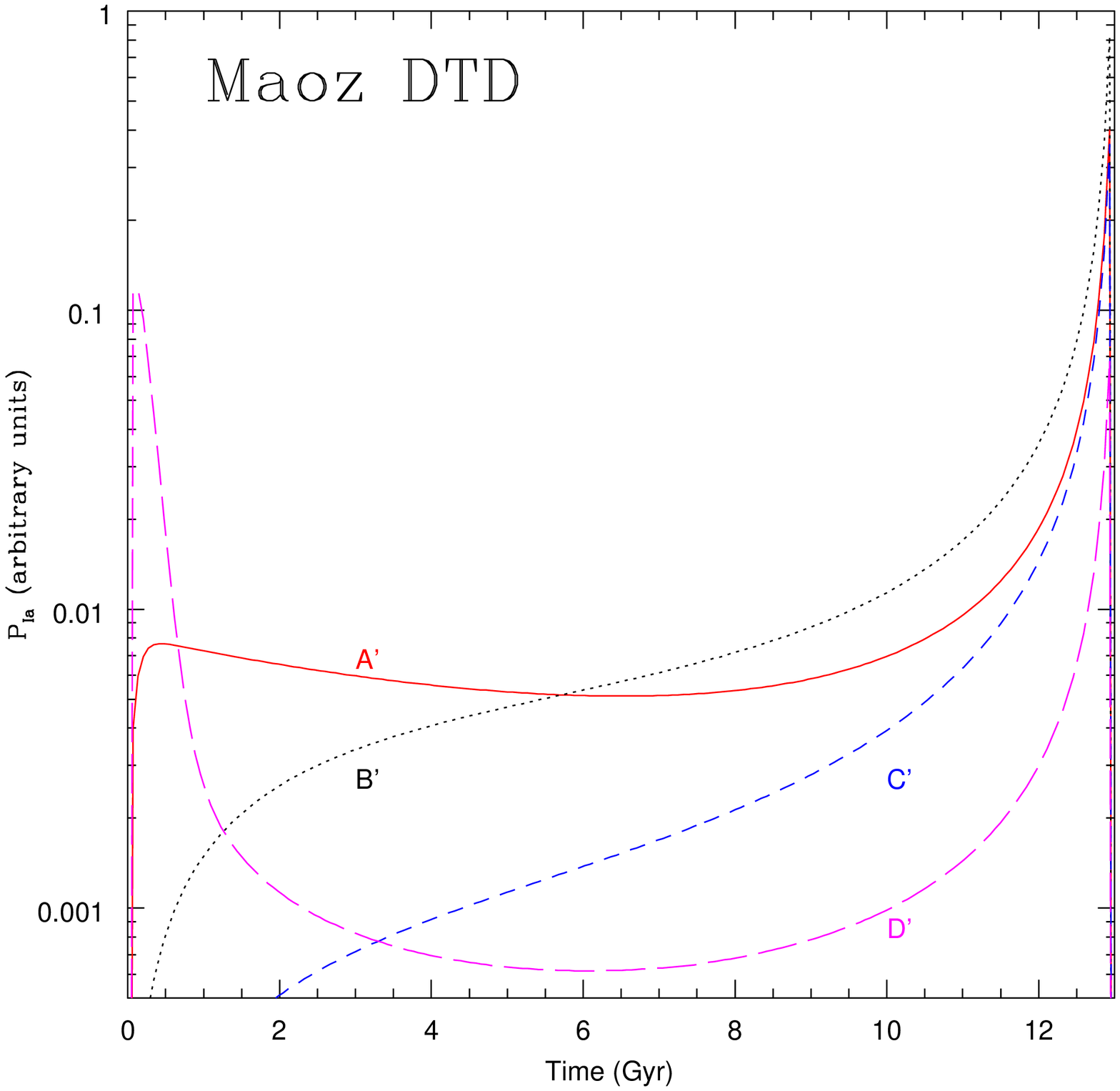}\\
   \includegraphics[width=8.8 cm]{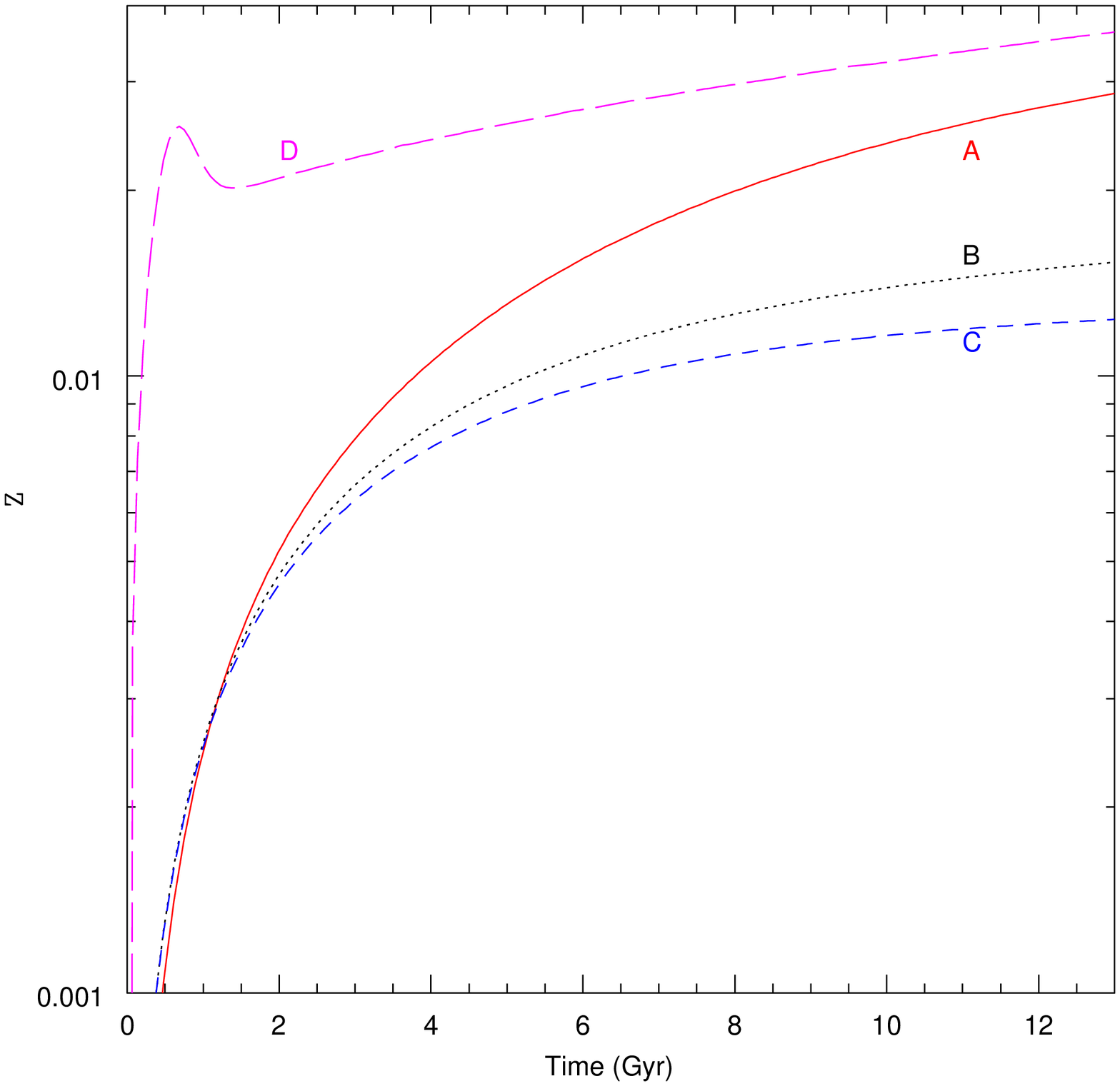}
   \includegraphics[width=8.8 cm]{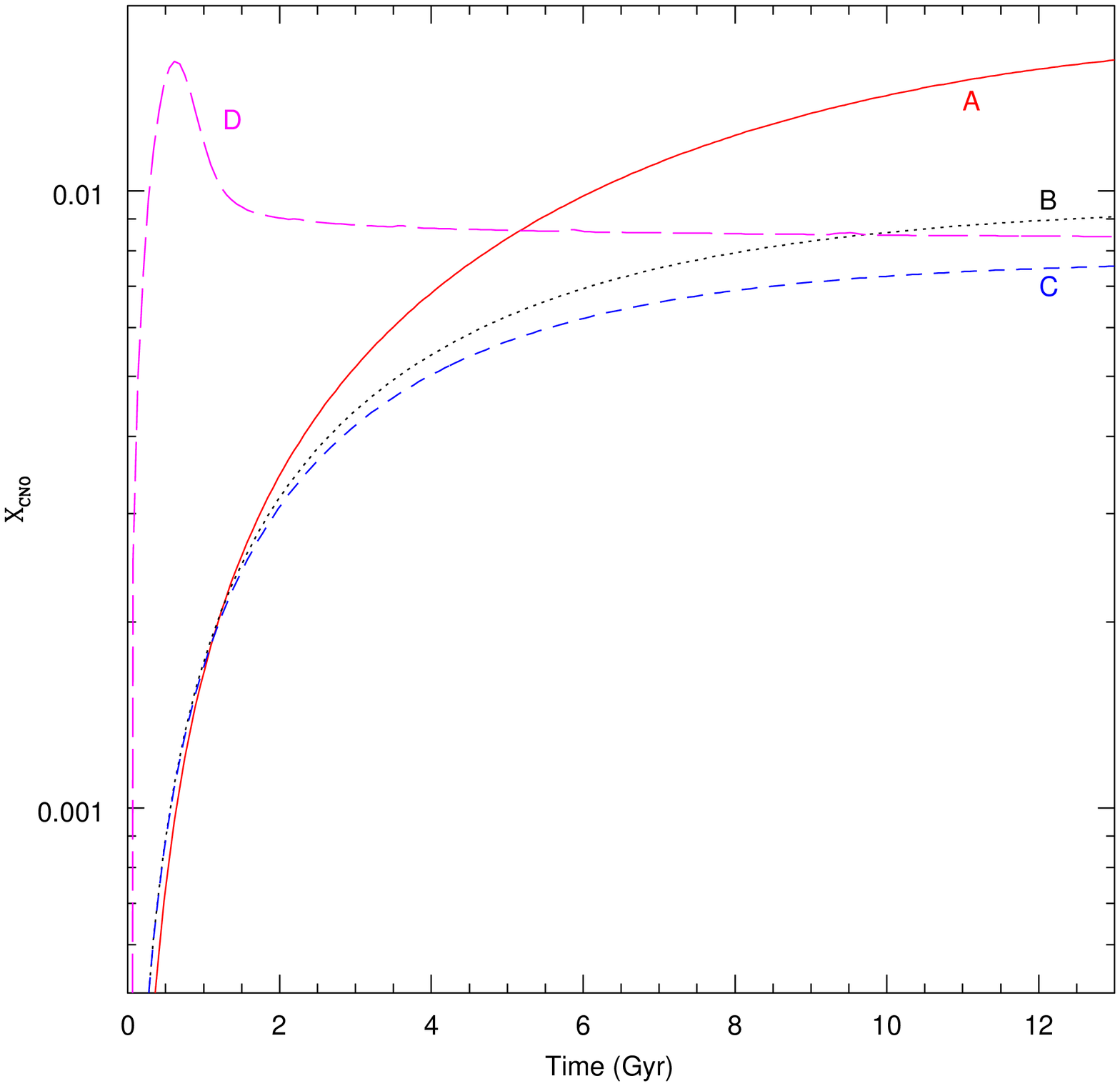}
\caption{Chemical evolution of the prototypical galaxies (the same models as in Fig.~\ref{figsfr}
and Table~\ref{tab1}). {\bf Top row:} Probability density that a SNIa exploding at $t=13$~Gyr was
born at a given cosmic time obtained with the DTD of P08 (left), and with the DTD of M10b (right).
The infall of gas that primes the formation of stars begins at a cosmic time $t=0$ for all the
models. {\bf Bottom row:} Evolution with time of the gas-phase metallicity (left), and the mass
fraction of CNO elements (right).
}\label{figpiaP}
\end{figure*}

In Fig.~\ref{figpiaP}, we show the distribution of birth-times for a SNIa exploding at
$t=t_\mathrm{gal}$. The curves peak at the present time because the DTD is strongly biased
towards small delay times. This effect is more pronounced in late-type galaxies, for which the
star-formation activity is either slightly decreasing or increasing. Early-type galaxies display a
similar peak at quite early times, belonging to the epoch of intense star-formation activity, but
the importance of this initial peak varies according to the duration of the star-formation epoch.
Thus, SNIa from early-type galaxy models come from two different populations whose properties
can be quite different: an old, predominantly low-mass, population belonging to the
star-formation peak, and a young population characterized by larger masses. 

\begin{figure}
\centering
   \includegraphics[width=9 cm]{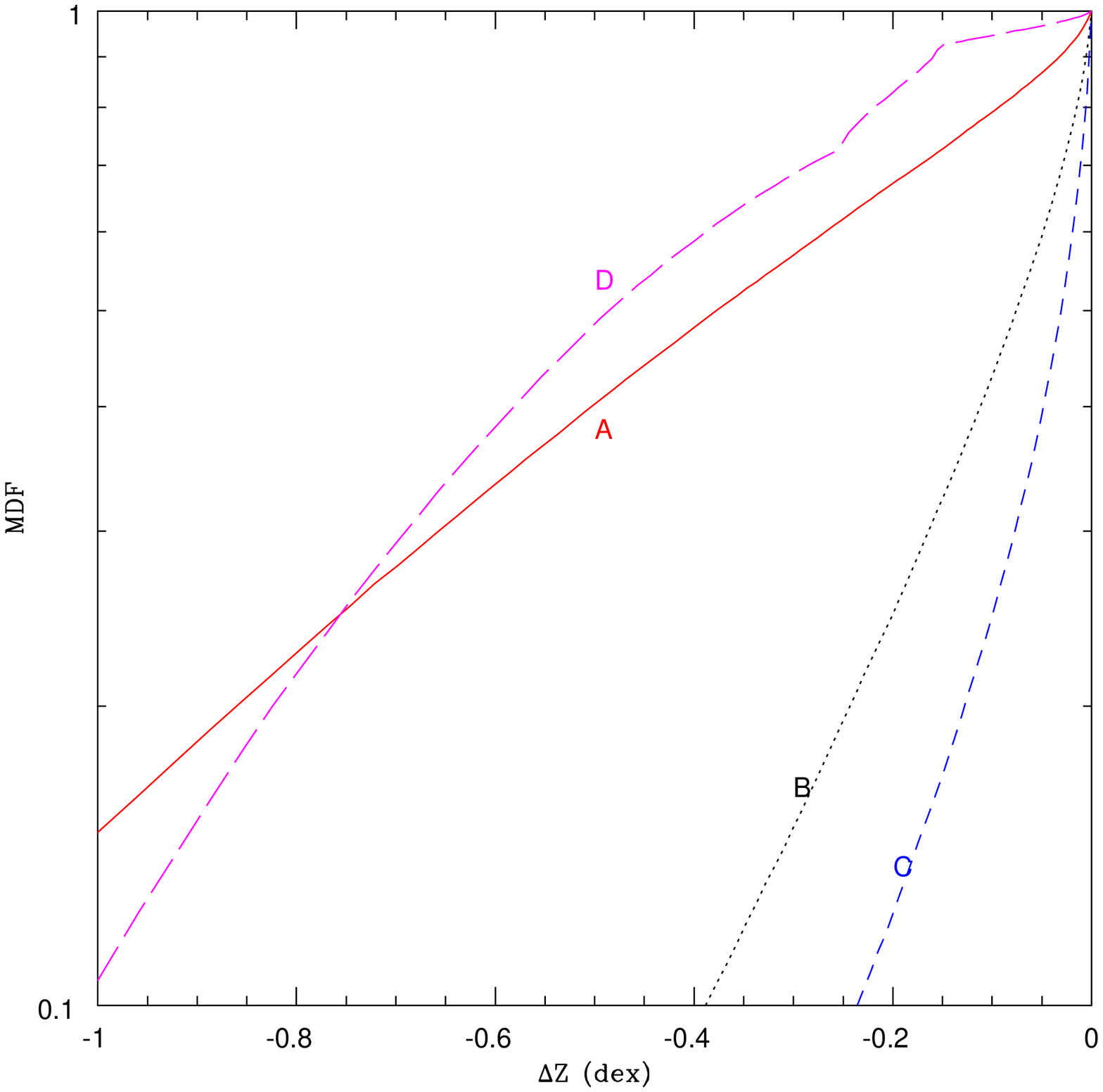}
\caption{
Metallicity distribution function (MDF) of SNIa as a function of the metallicity correction
with respect to $Z_\mathrm{host}$, for the same models as in the left panel of Fig.~\ref{figpiaP}.
}\label{figMDF}
\end{figure}

Figure~\ref{figMDF} shows the MDF of SNIa, for the same typical galaxies as those in the left
panel of Fig.~\ref{figpiaP}, as a function of $\Delta Z$.
As can be deduced from Fig.~\ref{figpiaP}, the younger the stellar population of a galaxy is (as
in models B and C) the smaller the scatter in metallicity of potential SNIa progenitors. 
The late-type galaxy distribution is strongly biased towards a zero correction and display a short
tail, down
to $\Delta Z\sim -0.2$~dex for a galaxy that is still in its initial phases of star formation
($\tau_\mathrm{inf}=50$~Gyr, meant to represent an irregular galaxy), and $\Delta Z\sim
-0.4$~dex for a galaxy that has already gone through most of the infall process
($\tau_\mathrm{inf}=9$~Gyr). The probability of a SNIa metallicity differing in less
than 0.1~dex (in absolute value) from $Z_\mathrm{host}$ is 75\% for the irregular galaxy model and
57\% for the model with $\tau_\mathrm{inf}=9$~Gyr.

Early-type galaxies show a broader MDF implying that SNIa from these hosts will have a larger
intrinsic metallicity dispersion, even if they come from the same host! Again, this behaviour can
be understood from the probability density of SNIa progenitors shown in Fig.~\ref{figpiaP}: Models
A and D have a substantial contribution of old potential SNIa progenitors, thus they show a large
scatter in their MDF. For the model with
$\tau_\mathrm{inf}=0.1$~Gyr and $\tau_\mathrm{sfr}=4$~Gyr half of the SNIa have a metallicity
differing more than 0.38~dex (in absolute value) from $Z_\mathrm{host}$, while for the model with
$\tau_\mathrm{inf}=0.1$~Gyr and $\tau_\mathrm{sfr}=0.1$~Gyr the median of the correction is
-0.49~dex.

\begin{figure}
\centering
   \includegraphics[width=9 cm]{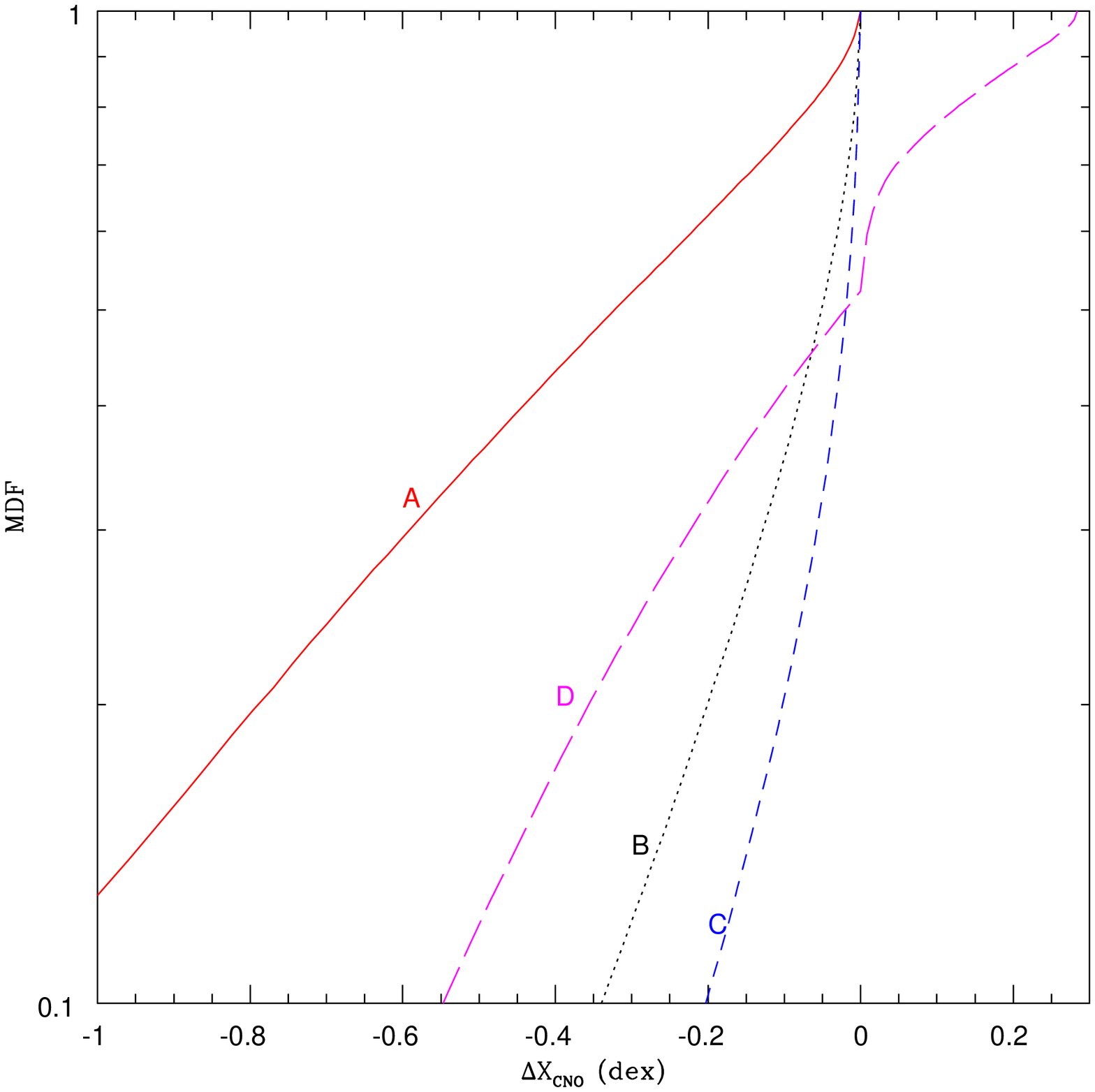}
\caption{
Metallicity distribution function (MDF) of SNIa as a function of the metallicity correction
with respect to the host metallicity, but taking the CNO abundance as a measure of metallicity.
The models shown are the same as in the left panel of Fig.~\ref{figpiaP}.
}\label{figCNOMDF}
\end{figure}

Figure~\ref{figCNOMDF} shows the MDF of SNIa when the metallicity is measured as the abundance of
CNO elements. The main difference with the MDF derived for 'Fe' is found in the
behaviour of the older galaxy model, i.e. the one with $\tau_\mathrm{inf}=0.1$~Gyr and
$\tau_\mathrm{sfr}=0.1$~Gyr. The reason of this behaviour can be understood by looking at
Fig.~\ref{figpiaP} (bottom right panel), where we display the evolution with time of the mass
fraction of CNO elements in the gas. The long-dashed (magenta) curve shows the evolution following
the short burst of star
formation coincident with the infall epoch. Initially the abundance of CNO elements grows fast due
to the self-enrichment of the gas due to efficient reprocessing driven by massive stars. Once the
star formation burst is over, there are no more massive stars present, and intermediate and low
mass stars enrich the interstellar medium at the end of their lifetime with yields representative
of their birth time, on average less rich on CNO elements than at the peak of the curve.
This causes the CNO abundance to decrease rapidly until it reaches a stable value at about 60\% of
the peak CNO mass fraction. Therefore, the SNIa that explode at $t=13$~Gyr can have birth
metallicites {\it larger} than that of the host, as seen in Fig.~\ref{figCNOMDF}. 

Because SNIa are observed at different redshifts, it is also interesting to know if their MDF
varies
with the elapsed time of galactic evolution, $t_\mathrm{gal}$. We have integrated again the
evolutionary equations with $t_\mathrm{gal}=11$ and 13.7~Gyr. Within this range of
$t_\mathrm{gal}$ the ``metallicity correction`` changes by less than $\sim10\%$.

\subsection{Statistical properties of the SNIa MDF for different types of galaxy}\label{sectstat}

In this section, we use a sample of galactic models with varying
$\tau_\mathrm{sfr}$, $\tau_\mathrm{inf}$, and $c_\mathrm{out}$ to study the statistical properties
of the relationship between $Z_\mathrm{Ia}$ and $Z_\mathrm{host}$. 
We have built models for early-type galaxies by using small values of the timescales
associated to infall and SFR: either $\tau_\mathrm{inf}=0.1$~Gyr
or $0.5$~Gyr, and either $\tau_\mathrm{sfr}=0.1$~Gyr or $0.5$~Gyr, and with different degrees of
outflow, controlled by the parameter $c_\mathrm{out}$. Thus, for each value of $c_\mathrm{out}$ we
show four models, obtained with the aforementioned combinations of $\tau_\mathrm{inf}$ and
$\tau_\mathrm{sfr}$. In general, the location of these models in the plane $\bar{Z}_\mathrm{Ia}$ vs
$Z_\mathrm{host}$ is driven by $c_\mathrm{out}$, while the variation with $\tau_\mathrm{inf}$ and
$\tau_\mathrm{sfr}$ is small. 

In order to generate models for late-type galaxies\footnote{In the
following we will refer as late-type galaxy models those models generated with
$\tau_\mathrm{sfr}\geq1$~Gyr and $c_\mathrm{out}=0$. Note that this includes models A to C and A'
to C' in Table~\ref{tab1}.} we left the infall and SFR timescales vary in
the ranges: $\tau_\mathrm{inf}=0.1 : 60$~Gyr and $\tau_\mathrm{sfr}=1 : 6$~Gyr. In order to avoid
too extreme or unrealistic galaxy models we have further constrained our sample to have $0.05\leq
f_\mathrm{gas}/f_\mathrm{star}\leq 3$ and $-0.6\leq \log\left(Z_\mathrm{host}/Z_\odot\right)\leq 0.5$.

\begin{figure*}
\centering
   \includegraphics[width=8.8 cm]{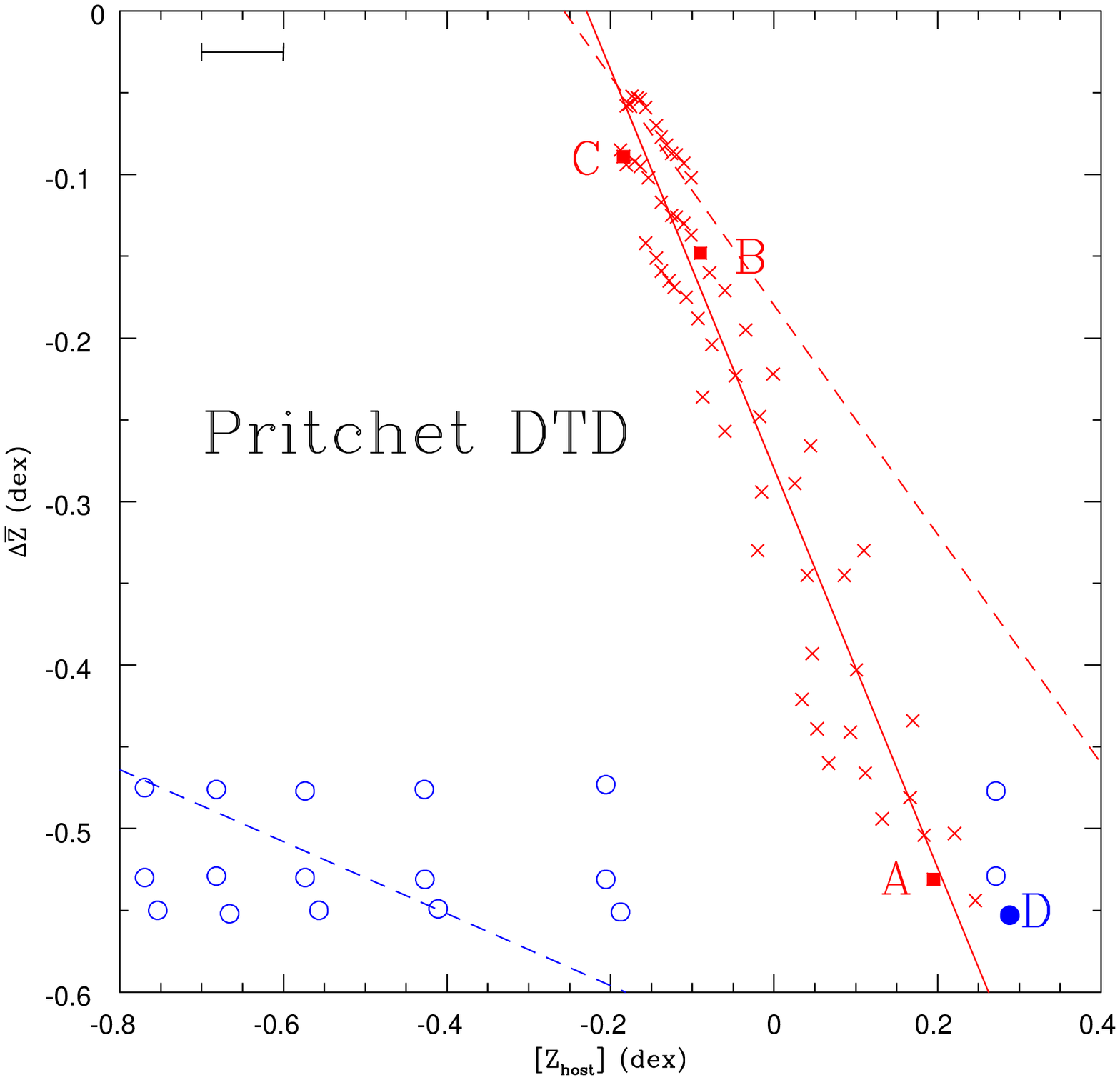}
   \includegraphics[width=8.8 cm]{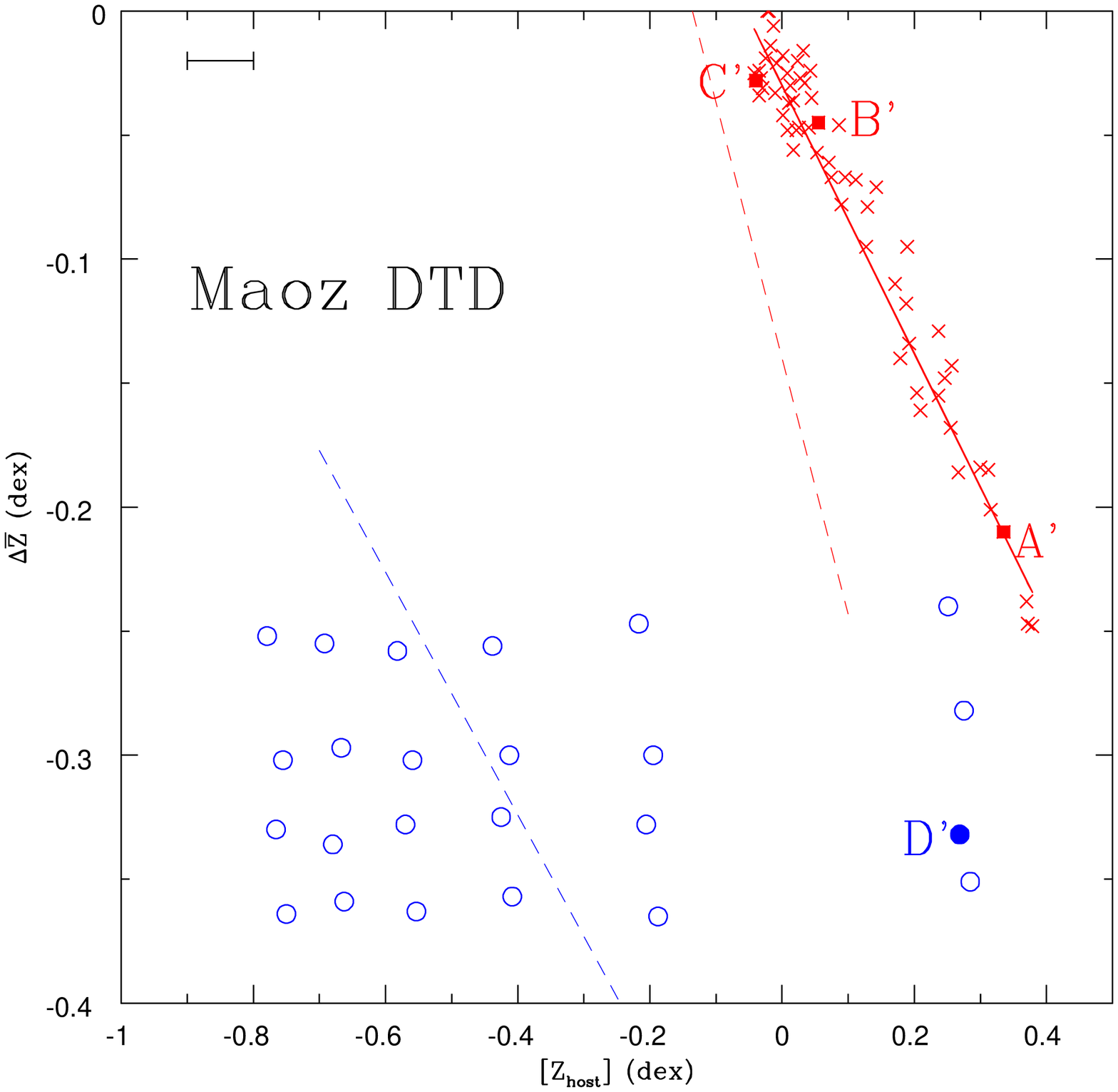}
\caption{
The metallicity correction, i.e. the mean difference between the logarithms of 
$\bar{Z}_\mathrm{Ia}$ and the host metallicity, $Z_\mathrm{host}$, as a function of the last. Blue
circles represent early-type galaxies, which we have
built with small values of the timescales associated to infall and SFR: $\tau_\mathrm{inf}=0.1$
and $0.5$~Gyr, and $\tau_\mathrm{sfr}=0.1$ and $0.5$~Gyr, and with different degrees of outflow,
controlled by the parameter $c_\mathrm{out}$. 
In this and coming figures we have plotted as late-type galaxies
(red crosses) all the models with $\tau_\mathrm{sfr}\geq1$~Gyr. 
The error bar at the top left of each panel is the typical uncertainty in observationally
determined $\left[Z_\mathrm{host}\right]$ (see Section~\ref{themostrelevant}).
{\bf Left:} Models computed with the DTD of P08.
The solid red line is a linear fit to the data points belonging to late-type galaxies given by:
$\Delta \bar{Z}=-1.22 \left[Z_\mathrm{host}\right] - 0.28$, where we have adopted a solar metal
fraction $Z_{\sun}=0.01895$. 
The dashed lines are linear fits belonging to early-type galaxies (blue) and late-type galaxies
(red) in the VESPA catalogue computed using the DTD of P08 (see Section~\ref{rescat}).
The filled symbols belong to models C, B, A, and D in Table~\ref{tab1} (from top to bottom).
{\bf Right:} Models computed with the DTD of M10b (note the different range of the vertical axis as compared to the left panel). 
The solid red line is a linear fit to the data points belonging to late-type galaxies (red
crosses), given by: $\Delta \bar{Z}=-0.54 \left[Z_\mathrm{host}\right] - 0.03$. 
The dashed lines are linear fits belonging to early-type galaxies (blue) and late-type galaxies
(red) in the VESPA catalogue computed using the DTD of M10b (see Section~\ref{rescat}).
The filled symbols belong to models C', B', A', and D' in Table~\ref{tab1} (from top to bottom).
}\label{figZiaP}
\end{figure*}

Figure~\ref{figZiaP} (left panel: DTD of P08) shows the metallicity correction, to be
applied to $Z_\mathrm{host}$ in order to estimate $Z_\mathrm{Ia}$, as a function of the host
metallicity. The two kind of galaxies show different behaviour:
while there is an approximately linear dependence of the correction on $Z_\mathrm{host}$ for late-type
galaxies, the correction is about constant for early-type galaxies of every metallicity, i.e. the dispersion of the corrections is much smaller than the range of host metallicities. The linear
function that fits the late-type galaxies passes as well through
the early-type ones with no outflow ($c_\mathrm{out}=0$). As $c_\mathrm{out}$ increases, the final
metallicities of the early-type hosts decrease while the metallicity correction remains the same. 

\begin{figure*}
\centering
   \includegraphics[width=8.8 cm]{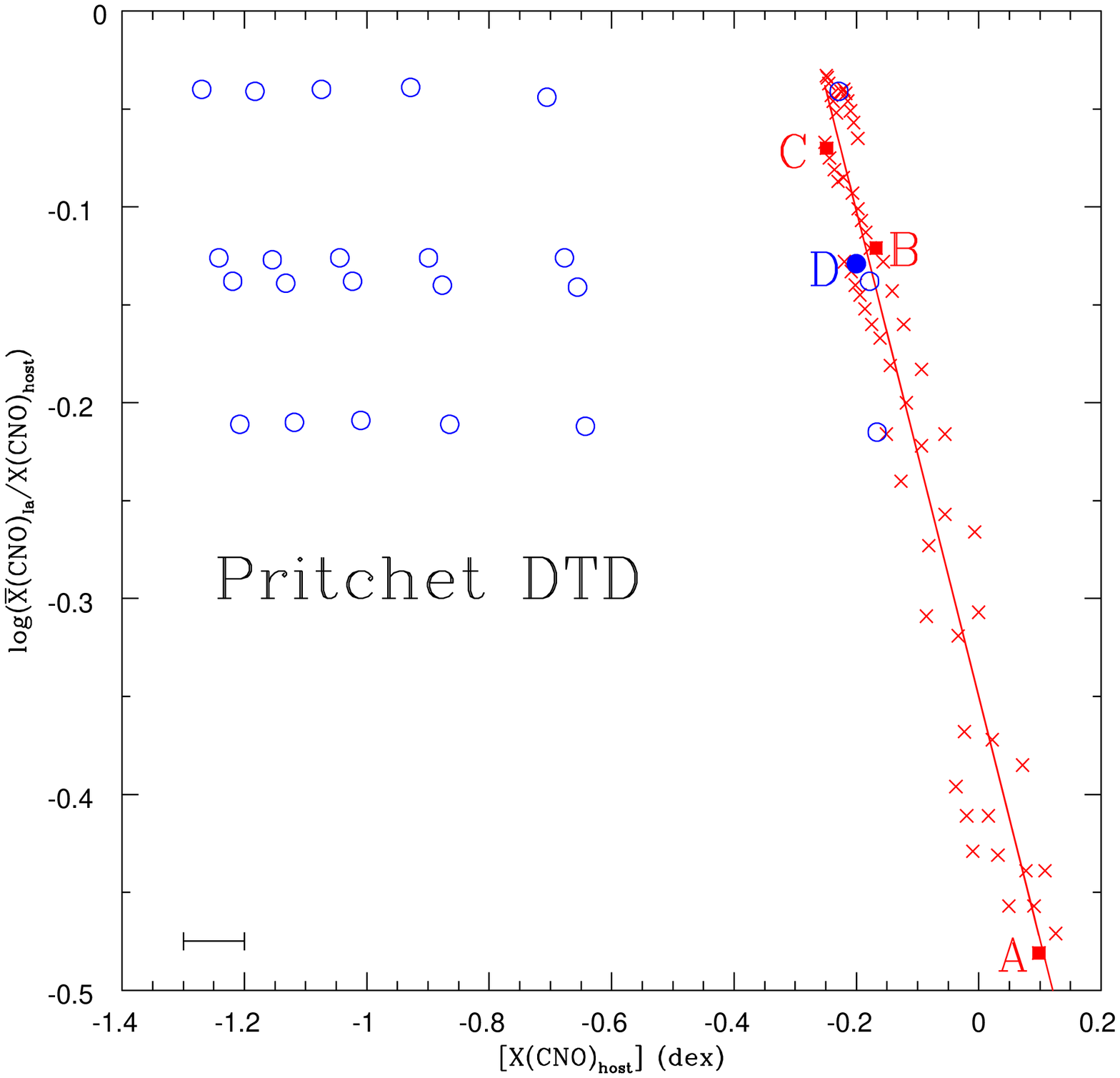}
   \includegraphics[width=8.8 cm]{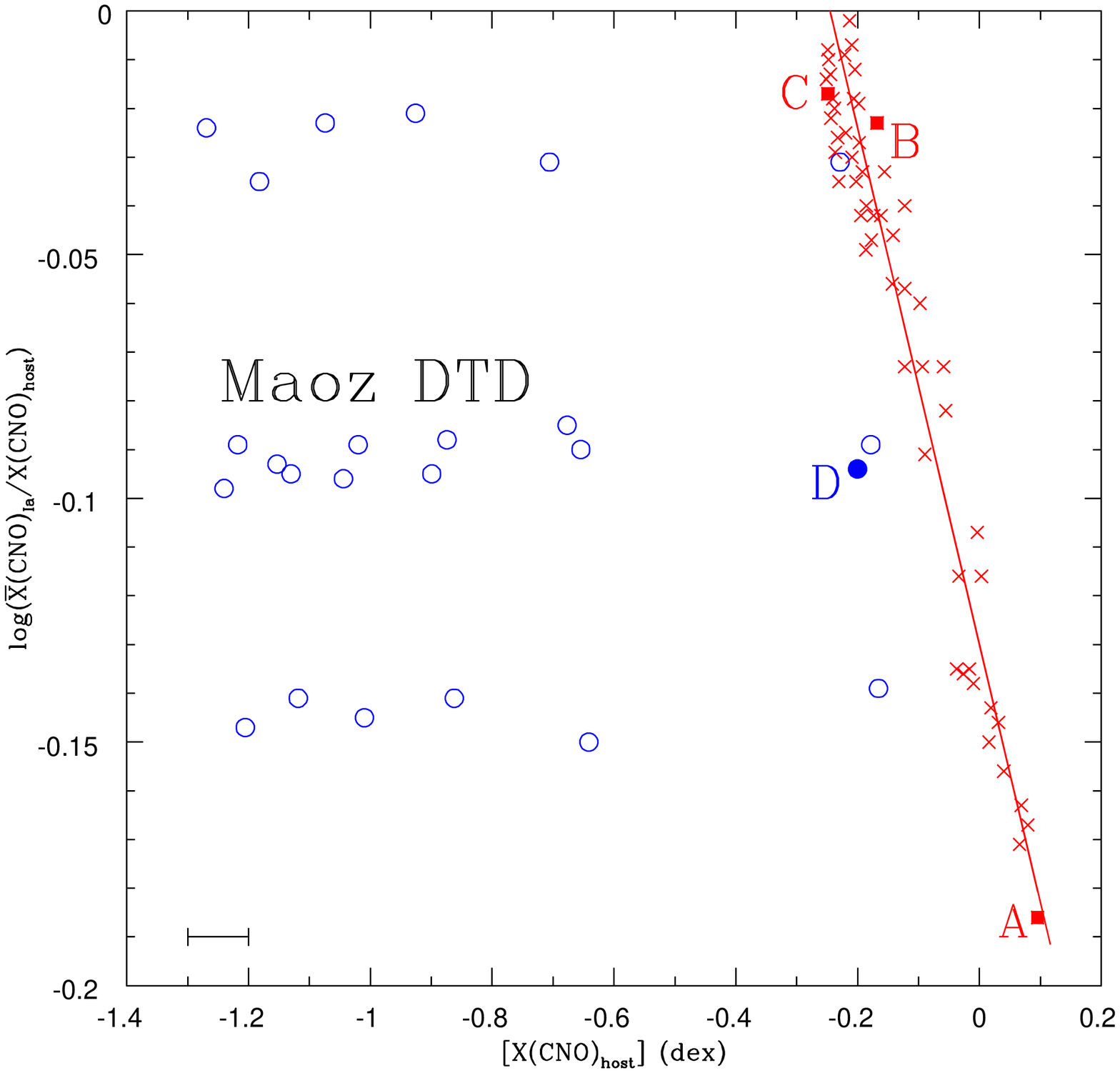}
\caption{
The metallicity correction with respect to the gas-phase metallicity of the host, when the
metallicity is measured as the CNO abundance. 
The error bar at the bottom left of each panel is the typical uncertainty in observationally
determined $\left[X(\mathrm{CNO})_\mathrm{host}\right]$ (see Section~\ref{themostrelevant}).
{\bf Left:} Models computed with the DTD of P08 (note the different range of the vertical axis as compared to the left panel). 
The straight line is a linear fit to
the data points belonging to late-type galaxies (red crosses), given by:
$\overline{\log}\left(X(\mathrm{CNO})_\mathrm{Ia}/X(\mathrm{CNO})_\mathrm{host}\right) = -1.24
\left[X(\mathrm{CNO})_\mathrm{host}\right] - 0.35$, where we have adopted a solar CNO mass
fraction $X(\mathrm{CNO})_{\sun}=0.0134$. From top to bottom, the filled circles
belong to models C, B, D, and A in Table~\ref{tab1}. 
{\bf Right:} Models computed with the DTD of M10b. The straight line is a linear fit to
the data points belonging to late-type galaxies (red crosses), given by:
$\overline{\log}\left(X(\mathrm{CNO})_\mathrm{Ia}/X(\mathrm{CNO})_\mathrm{host}\right) = -0.53
\left[X(\mathrm{CNO})_\mathrm{host}\right] - 0.19$. From top to bottom, the filled circles belong
to models C', B', D', and A' in Table~\ref{tab1}. 
}\label{figcnoiaP}
\end{figure*}

Figure~\ref{figcnoiaP} shows the metallicity correction belonging to CNO abundances. It
displays the same gross properties as those seen in Fig.~\ref{figZiaP} with one important
difference concerning early-type galaxies: their metallicity corrections are now among the lowest of all
models displayed. The reason for this behaviour is again the peculiar MDF of CNO
elements in SNIa from this kind of galaxies as seen in Fig.~\ref{figCNOMDF}. The dispersion of
their MDF is, however, similar to that of galaxies with a much larger metallicity correction
(compare models A and D in Table~\ref{tab1}).

In many observational studies, it is necessary to generate a random set of SNIa with stochastic
metallicities in order to compare their statistical properties with those of the observational
sample. The linear fit of $\Delta\bar{Z}$ vs. $[Z_\mathrm{host}]$ shown in Fig.~\ref{figZiaP}
together with the MDF may be used to generate such a random set of SNIa, with stochastic
metallicities related (but not equal) to their host metallicities. The simpler way to do this is
to approximate the MDF of SNIa as a straight line in the semi-logarithmic plot
shown in Fig.~\ref{figMDF},

\begin{equation}
\log (\mathrm{MDF}) = \alpha\Delta Z\,,
\label{eqfitMDF}
\end{equation}

\noindent that determines the probability density,

\begin{equation}
 P'(\Delta Z) = \alpha\exp\left(\alpha\Delta Z\right)\,,
\end{equation}

\noindent the mean metallicity correction,

\begin{equation}
\Delta \bar{Z} = -1/\alpha\,, 
\end{equation}

\noindent and the standard deviation, $\sigma$, of the distribution of $\Delta Z$: 

\begin{equation}
 \sigma = 1/\alpha = -\Delta \bar{Z}\,.
\label{eqsigma}
\end{equation}

\noindent In Table~\ref{tab1} we give the standard deviation of the MDF of our typical galaxy
models. As can be seen, the relationship between $\Delta\bar{Z}$ and $\sigma$ derived in
Eq.~\ref{eqsigma} approximately holds for all the models, with the exception of the distribution of
CNO in models D and D', for which a simple proportionality law as that of Eq.~\ref{eqfitMDF} is
not a good fit (see Fig.~\ref{figCNOMDF}).

To generate a stochastic distribution of SNIa metallicites in a late-type galaxy of known
$Z_\mathrm{host}$, one would have first to apply the linear function of Fig.~\ref{figZiaP} (or
that in Fig.~\ref{figcnoiaP} if what is desired is the distribution of mass fractions of the CNO
elements) to obtain the metallicity correction $\Delta \bar{Z}$ and
$\sigma$ ($=-\Delta \bar{Z}$). For an early-type host the procedure is even
simpler, as the metallicity correction takes on a nearly constant value for this kind of galaxies.

\subsection{Choice of the Delay Time Distribution}\label{sectdtd}

The Delay Time Distribution is a property specific of SNIa that has rather
small imprint on the galactic chemical evolution of CNO elements but affects strongly the temporal
distribution of SNIa, hence it is expected to influence
appreciably the MDF. The DTD of SNIa is a matter of current debate
\citep{stro04,sca05,gre05,gre08,mao10}. P08 proposed a model in which the rate 
of SNIa was proportional to the WD formation rate. This model 
reproduced satisfactorily the trend of SNIa rates as a function of the specific SFR in galaxies of
ages between 1 and 13~Gyr. In a recent work, \citet{mao10} presented a method
to recover the supernova DTD that simultaneously takes into account supernova data and the
reconstructed star formation history of the individual galaxies in a survey, and applied this
method to the events present in the LOSS SN survey and in the VESPA database. 
Independently, \citet{brandt10} analyzed light curves and host galaxy spectra of 101 SNIa from the 
SDSS using VESPA algorithms, from which they derived the DTD of SNIa.
\citet{mao10} found
that their data required a stronger short delay-time component than allowed by the model of
P08, i.e. they called for a higher contribution from large mass progenitors. The VESPA
database is organized in temporal bins, whose resolution is better at the present epoch and
degrades with look-back time (for further details see Section~\ref{catalogs}), hence the results of
\citet{mao10} should be more sensitive to the short delay time component of the SNIa DTD. Thus, the
discrepancy between the models proposed by P08 and \citet{mao10} might be due to the
different sensitivities to the time ranges of the SNIa DTD (but see M10b). Anyway, it is
necessary to check the
imprint that the different DTD have on the MDF of SNIa. 

Having used a model based on P08 in the previous sections, we now discuss the results
obtained with the DTD of M10b, Eq. 13 in their paper,
\begin{equation}
 \Psi(t) = 0.7\times10^{-3} t^{s}\, \mathrm{SN}~\mathrm{M}_\odot^{-1}~\mathrm{Gyr}^{-1}\,,
\label{DTDMaoz}
\end{equation}
\noindent where $s=-1.2$, $t$ is measured in Gyr, and we have applied a factor 0.7 to the DTD of
M10b to convert back from their 'diet-Salpeter' IMF to ours. 
Models A' to D' in Table~\ref{tab1} were computed using their DTD in Eqs.~\ref{eqria} and
\ref{eqpia}. The right panel of 
Fig.~\ref{figpiaP} shows the distribution of birth-times for a SNIa exploding at $t=t_\mathrm{gal}$
for these model galaxies. With respect to the DTD of P08 (left panel of 
Fig.~\ref{figpiaP}) the new DTD increases the contribution from young stars (small delay-time)
with respect to old stars. Late-type galaxies are scarcely affected at all, but early-type
galaxies display a quite different trend in which the contribution of the population belonging to
the initial star-formation peak is dramatically reduced. With this new DTD, the differences
between galaxy-types strongly smooth out. 

The metallicity correction obtained with the DTD of M10b is shown in the right panel of
Figs.~\ref{figZiaP} and \ref{figcnoiaP}, for both late-type and early-type galaxies. Overall, the
correction is significantly smaller than that obtained using the DTD of P08, as expected given the
larger contribution from prompt SNIa implied by the DTD of M10b. 

From these results it is clear that the DTD of SNIa is the most important factor determining the
magnitude of the correction needed to obtain the metallicity of SNIa from those of their host
galaxies. We stress that any of the above DTD models (P08 vs. M10b) is subject to a
large degree of observational uncertainty. For instance, M10b discuss two possible approaches to
determine the power law exponent in Eq.~\ref{DTDMaoz}, based on what they call the 'optimal-iron
constraint' and the 'minimal-iron constraint', which lead to different exponents, $s$, in the range
from -1.5 to -0.9 (in this study, we have adopted in Eq.~\ref{DTDMaoz} the mean, i.e. -1.2). It is
interesting to consider how much would change the plots in Figs.~~\ref{figZiaP} and \ref{figcnoiaP}
if $s$ covered the
full range given by M10b. We have repeated the calculations with $s=-1.5$ and
$s=-0.9$. For the first case, belonging to an extreme case of the 'optimal-iron constraint', the
straight-line fit to
our late-type galaxies changes to $\Delta\bar{Z}=-0.28\left[Z_\mathrm{host}\right] + 0.06$ while
the correction for early-type galaxies changes by $\sim+0.12$~dex. For the last case, belonging
to an extreme case of the 'minimal-iron constraint', the fit of the late-type host corrections
changes to $\Delta
\bar{Z}=-0.89\left[Z_\mathrm{host}\right] - 0.16$, and the corrections for early-type hosts
changes by $\sim-0.04$~dex. We can see that the uncertainty in the DTD leads to uncertainties in
$\Delta\bar{Z}$ of the same order as the own correction. We conclude that an accurate
determination of the DTD of SNIa is a prerequisite to achieve a good understanding of the
differences between the metallicities of these supernovae and those of their host galaxies.

\subsection{Is it the host metallicity a good estimator of the SNIa
metallicity?}\label{themostrelevant}

Here we discuss the reliability of different host metallicity estimators as for establishing the
statistical properties of SNIa: gas-phase vs. stellar metallicity, $Z_\mathrm{star}$, and gaseous
CNO abundance vs. Fe\footnote{Recall that what we are calling here 'Fe' represents all the metals.}
abundance.

We have evaluated the mean host stellar metallicity as the mean of the metallicity of the ISM at
star birth, weighted by the SFR and by the IMF, for all the stars whose mean-sequence lifetime is
larger than the elapsed time since their formation. The mean stellar metallicity derived
by observational studies is weighted by the contribution of stars in different phases of their
life, with weights depending on the particular spectral feature used for measuring the metallicity.
Even though both ways to define a mean stellar metallicity may give different values, we expect
that both will follow the same qualitative trends.

\begin{figure}
\centering
   \includegraphics[width=8.8 cm]{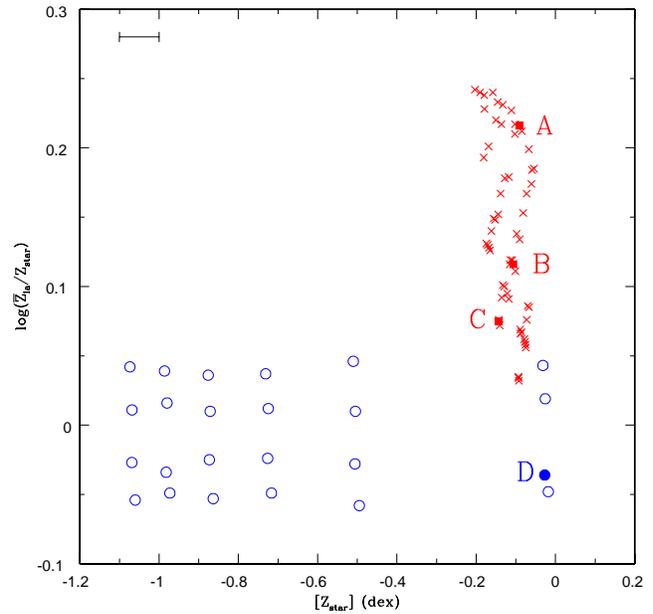}
\caption{
The metallicity correction, taking as a reference the mean stellar metallicity,
$Z_\mathrm{star,}$, instead of the gas metallicity, $Z_\mathrm{host}$. Here we have used the DTD of
M10b. 
The filled symbols belong to models A',
B', C', and D' in Table~\ref{tab1} (from top to bottom).
The error bar at the top left of the figure is the typical uncertainty in observationally
determined $\left[Z_\mathrm{star}\right]$ (see discussion at the end of 
Section~\ref{themostrelevant}).
}\label{figZiaZstar}
\end{figure}

Fig.~\ref{figZiaZstar} shows the metallicity correction to apply if the host
galaxy metallicity is measured from the stellar population (and the DTD of M10b is used). In
comparison with the trend shown in the right panel of 
Fig.~\ref{figZiaP} we notice that the corrections to apply to the metallicities of early-type
galaxies continues being independent of host metallicity, but now they cluster around
$\Delta\bar{Z}=0$. Late-type
galaxies display also a similar behaviour as in Fig.~\ref{figZiaP}, with changed sign, but they
could not be fitted
by a linear function and show larger dispersion for given $Z_\mathrm{host}$. Thus, if possible, it
is advisable to measure the metallicity of SNIa in late-type hosts as that of their
gas.
These results can be understood by computing the distribution of birth-times of present-day (at
$t=t_\mathrm{gal}=13$~Gyr) stars. We have calculated the 
distribution of birth times as the fraction of the mass in present-day stars
represented by the stars born at time $t$ that are still alive,
\begin{equation}
 \frac{\mathrm{d}m_\mathrm{star}}{\mathrm{d}t} = S(t)\int_{M_\mathrm{bot}}^{M(t_\mathrm{gal}-t)}
\phi(M) M
\mathrm{d}M\,.
\label{birth}
\end{equation}
The distribution of birth-times follows closely the SFR of each model (see Fig.~\ref{figsfr})
because 99\% of the stellar mass at birth belongs to stars smaller than $\sim0.5$~M$_\odot$ (for
our Salpeter IMF), hence the integral in Eq.~\ref{birth} is always close to $\sim1$. 
Comparing Fig.~\ref{figsfr} with the probability density of birth times of SNIa shown in
Fig.~\ref{figpiaP}
(top right) it is easy to see that the average age of SNIa progenitors is much less than the
average age of the stellar population, especially because the peak of $P_\mathrm{Ia}$ at
times close to $t_\mathrm{gal}$. In galaxies characterized by a monotonic increase of the
metallicity (models
A', B', and C') the immediate consequence is that the metallicity of SNIa is larger than the
average stellar metallicity. Early-type galaxies, like model D', display a non-monotonic evolution
of the metallicity, that peaks later than the SFR. For instance, for model D' the peak of the SFR
occurs at $t\simeq0.1$~Gyr, while the peak of $Z_\mathrm{gas}$ is found at $\sim0.85$~Gyr and that
of $X_\mathrm{CNO}$ takes place at $\sim0.60$~Gyr. Thus, the stellar population is chiefly drawn
from low-metallicity gas, while SNIa have an appreciable contribution from times close to
present, at which the gas metallicity is much larger. 

In Fig.~\ref{figcnoiaP} we show the metallicity correction when the gas-phase metallicity is
estimated from the mass fraction of CNO elements in the gas. In comparison with
Fig.~\ref{figZiaP} the CNO metallicity correction is nearly identical to the Fe correction for
late-type galaxies but significantly smaller for early-type hosts, for a given DTD. 
As can be seen in Fig.~\ref{figpiaP} (bottom row) the chemical evolutions of Fe and CNO elements
in the galaxy models A', B', and C' are in close agreement, thus the MDF of both element groups
are similar (compare Fig.~\ref{figMDF} with Fig.~\ref{figCNOMDF}).
Thus, the use of Fe instead of CNO abundances does not seem to represent an improvement in the
estimates of SNIa metallicities obtained from those of their hosts.
The host CNO abundance of early-type galaxies is in much better agreement with the {\it mean} SNIa
one than are the corresponding total metal abundances, although their dispersions are
similar. Using again the chemical evolution computed for the prototypical galaxy model D',
displayed in the bottom row of Fig.~\ref{figpiaP}, and the probability density of SNIa shown in
Fig.~\ref{figpiaP}, the reason for this closer agreement between CNO abundances in SNIa and the
gas, as compared with Fe abundances, is that SNIa from early-type galaxies have non-negligible
contributions from quite early times and near present time. The CNO abundances are low at both
times, because presently there are no high-mass stars contributing to their synthesis. On the
other hand, the Fe abundance increases continuously during the last 11 Gyr due to the contribution
of SNIa, with the result that young SNIa progenitors are drawn from gas with high Fe abundance
while old SNIa progenitors were born from gas with quite small Fe abundance. Hence, the average
discrepancy between present-day gas Fe abundance and that of SNIa is larger than in the case of the
CNO abundances.

A meaningful evaluation of the corrections we propose to obtain SNIa metallicities from the
different metallicity measures of their hosts can only be done by considering the observational
errors in the galactic metallicity and in the DTD of SNIa. As discussed in the Introduction,
galactic metallicities are measured using different methods, all of them subject to a suite of
experimental and theoretical uncertainties. \citet{gal08} measured the Fe abundance of the gas of
early-type hosts of SNIa with quite different errors, depending on the host. An average error in
$\left[Z_\mathrm{host}\right]$ of 0.1-0.2~dex can be deduced from their Table~4 and from the
comparison they perform with metallicities measured by \citet{tra00} and \citet{tho05} for a subset
of galaxies they have in common. On the other hand, the \citet{tre04} mass-metallicity
relationship, used by \citet{how09} to measure host metallicity, is sensitive to
$\left[\mathrm{O}/\mathrm{H}\right]$ in the gas phase. The scatter in this relationship is of
order $\sim0.1$~dex. Finally, the observational determination of $\left[Z_\mathrm{star}\right]$ is
subject to systematic biases of order $0.1-0.2$~dex \citep{cid05}. Summarizing, the uncertainty in
the measured metallicity of SNIa hosts is similar for the different metallicity tracers,
$\epsilon\sim0.1$~dex, although it may be substantially larger for a given galaxy. 

We have drawn in Figs.~\ref{figZiaP}, \ref{figcnoiaP}, and \ref{figZiaZstar} an error bar
indicative of the typical uncertainty in the measured host metallicity. These uncertainties are
not large compared to the range of host metallicities covered in the figures (horizontal axes). One
can also
compare the host metallicity uncertainty with the correction necessary to obtain $Z_\mathrm{Ia}$,
plotted in the same figures (vertical axes). Looking at the corrections deduced for the DTD of P08
(left panels of
the same figures) it can be seen that they are in general much larger than $\epsilon$. On the
contrary, if the DTD of M10b is used (right panels of Figs.~\ref{figZiaP} and \ref{figcnoiaP}),
$\Delta\bar{Z}_\mathrm{Ia}$ is of the same order as $\epsilon$, especially for early-type galaxies.
Thus, if this last DTD were the
correct choice, the error in the determination of SNIa metallicity would be dominated by the
uncertainty in the measurement of the host metallicity rather than the difference between galactic
and supernova metallicities. 

\section{Exploiting galactic databases}\label{catalogs}

In this section, we explore the facilities provided by current galactic catalogues in order to
obtain observational constraints to our theoretically derived MDF of SNIa.
Ideally, one might use an observationally determined star formation history (SFH) of a galaxy
together with a gas metallicity history, determined through metallicity-dependent
synthetic stellar population (SSP) models, to recover a probability distribution function for the
metallicity of SNIa progenitors. In practice, however, there are limitations due to both the
discretisation of the catalogue in time bins of finite
size, and the paucity of metallicities used in the SSP models. In Section~\ref{probcat} we shortly
address how to reformulate Eqs.~\ref{eqpia2} and \ref{eqpia3} in order to calculate the probability
that a SNIa progenitor was born in a given time bin. We apply then the new formalism to the VESPA
catalogue, and discuss the results and the consequences of the paucity of metallicities in the SSP
models in Section~\ref{rescat}. We explain the main characteristics of the VESPA catalogue in the
next paragraph.

VESPA \citep['VErsatile SPectral Analysis',][]{toj07} is a method to reconstruct 
star formation and metallicity histories from galactic spectra using SSP models. \citet{toj09}
applied their VESPA code to all galaxy spectra in the seventh data
release of the SDSS \citep{yor00}, and compiled their results in a publicly
accessible catalogue of stellar masses, star formation rates and metallicity histories of nearly
800,000
galaxies \footnote{http://www-wfau.roe.ac.uk/vespa/index.html}. VESPA uses all of the available
absorption features and the shape of the continuum (emission lines are not included in the
analysis) in order to estimate the SFH of each galaxy, with variable time resolution, and the
metallicity of the stars born in each time bin, which traces the gas metallicity. 
The number of time bins depends on the quality of the data on each galaxy. At the highest
resolution, VESPA uses 16 age bins, logarithmically spaced between look-back times
0.002~Gyr and 13.7~Gyr, however if the data are not of sufficient quality some bins may lack
information. The VESPA catalogue contains results from several runs using different SSP models, IMF
prescriptions, dust models, and galactic samples (for details see the above-mentioned papers from
Tojeiro et al).
The SSP models are supplied at metallicities of $Z = 0.0004$, 0.01, 0.02, and 0.04.

\subsection{Probability that a SNIa originated in a given time interval}\label{probcat}

The present rate of SNIa explosions in a given host is the convolution over time of the
SFR and the DTD. The probability that a SNIa that explodes now (at
time $t$) comes from a progenitor born at time $t' = t - \tau$ is given by Eq.~\ref{eqpia}.
Assuming we know the DTD and that the SFR can be obtained from a galactic catalogue, both in
discrete time bins, the probability that the SNIa was born at time bin $i$ is (for a given galaxy):

\begin{equation}
 P_i = \frac{\Psi_i \mathcal{M}_i}{\sum \Psi_j \mathcal{M}_j}\,,
\end{equation}

\noindent where the addition in the denominator extends to all bins defined for that galaxy, and
$\mathcal{M}_i$ is the mass of stars formed in time bin $i$. 

If the observational DTD is not known with enough resolution (as compared to the time bin
resolution in a given catalogue), one can turn to the model of P08 (Eq.~\ref{eqpia2})
or its equivalent based on the DTD of M10b (Eq.~\ref{eqpia3}).
What we actually need in order to use the galactic catalogue is the probability that a
SNIa exploding now comes from a progenitor born within a time interval given by the temporal
bins defined in the catalogue: say between look-back times $\tau_i$ and $\tau_i+\Delta t_i$:

\begin{equation}
 P_i \propto \int_{\tau_i}^{\tau_i+\Delta t_i} S(t-t') \phi(M_{t'}) 
\left(\frac{dM}{dt}\right)_{t'} dt'\,,
\end{equation}

\noindent where we have used the DTD of P08.
To use the catalogued data we transform the above integral using a mean SFR defined from
the stellar mass formed in the time interval $i$ and the duration of that time
interval,
$\Delta t_i$: $\bar{S}_i = \mathcal{M}_i / \Delta t_i$, and combine $(dM/dt) dt' = dM$:

\begin{equation}
 P_i \propto \frac{\mathcal{M}_i}{\Delta t_i} \int_{M(\tau_i+\Delta t_i)}^{M(\tau_i)} \phi(M) dM\,.
\end{equation}

\noindent The integral in the last equation can be easily computed given the 
IMF, the stellar lifetime law, and the temporal limits of each bin. Note that there is no need to
know the proportionality constant for $P_i$ nor even to re-normalize the IMF, because we
know that $\sum P_i = 1$, if we add $P_i$ for all the bins in a galaxy. Thus, the probability
density can be normalized unambiguously. 

If the DTD of M10b is adopted instead, $P_i$ can be obtained from:

\begin{equation}
 P_i \propto \frac{\mathcal{M}_i}{\Delta t_i} \bar{\Psi}\,,
\label{eqpi}
\end{equation}

\noindent where $\bar{\Psi}$ is the average DTD between $\tau_i$ and $\tau_i+\Delta t_i$.

 From the previously determined probability that
a SNIa comes from time bin $i$, $P_i$, and the catalogued metallicity for that same time bin,
$Z_i$, the metallicity probability density is given by:  

\begin{equation}
 P'(\Delta Z) = P'(\log Z_i - \log Z_{host}) = P_i \,.\label{eqppi}
\end{equation}

\noindent This procedure provides us with a discrete set of $\Delta Z$ for which the
probability density is known.

\subsection{Results using VESPA: dependence on the star formation rate and host age}\label{rescat}

Here we show an example of the application of the formalism introduced in the last Section to
real data. We have accessed the VESPA catalogue (runID=2) and classified the galaxies as early or
late-type according to the criterion proposed by \citet{dil10} on base of
their $u$ and $r$ SDSS model magnitudes. Then, we have selected the galaxies with a minimum
temporal resolution of three bins, in order to work with reasonable SFHs. Finally, we recovered
90\,906 early-type galaxies and 40\,419 late-types.
We stress that we have not filtered
the catalogue in order to select which galaxies are potential SNIa hosts. From a theoretical point
of view, every single galaxy can house binaries with the appropriate parameters (total mass,
secondary mass, initial separation, metallicity) to produce SNIa. Observationally it is another
story, because surveys aiming to detect SNIa have to pre-select target galaxies for which the
observations and measurements are more efficient. This difference has to been kept in mind when
comparing between our theoretically derived SNIa statistics and the correlations found in
observational campaigns.

We have applied the formalism developed in the previous section (Eqs.~\ref{eqpi} and \ref{eqppi})
to all the galaxies selected from
VESPA in order to obtain the MDF of each galaxy. Unfortunately, at the present level of resolution
of the catalogue the resulting MDF is of little practical use because of the limited spectra of
metallicities
of the SSP models. As the MDF is an statistical description of the difference of metallicity
between the host and the SNIa, the use of just four metallicities in the SSP models produces an
unreliable extremely clumpy MDF. 
Keeping this in mind, we comment here briefly on the relationship between the metallicity
correction, $\Delta\bar{Z}$, and the $Z_\mathrm{host}$ obtained from the catalogue. Linear fits 
to these quantities are shown as dashed lines in Fig.~\ref{figZiaP}. Given the uncertainties, it is
striking the close coincidence between the points belonging to our theoretical models and the
linear fits obtained with the catalogue, especially for late-type
galaxies. The differences between theoretical models of early-type and late-type galaxies are more
or less reproduced by the mean metallicities derived from VESPA. 

In view of the strong uncertainty affecting the metallicity statistics, we do not follow further
with
the analysis of the VESPA derived MDF, but we emphasize
that the formalism presented in the previous subsection can (and should) be applied to future
galactic catalogues based on improved data and metallicity resolution. 
In spite of the failure to derive useful MDF from VESPA, the data from the catalogue can still be
used to show trends of the mean SNIa metallicity, $\bar{Z}_\mathrm{Ia}$ calculated using the
probability defined by Eq.~\ref{eqpi}, vs. the galactic SFR and age. 

\begin{figure}
\centering
   \includegraphics[width=9 cm]{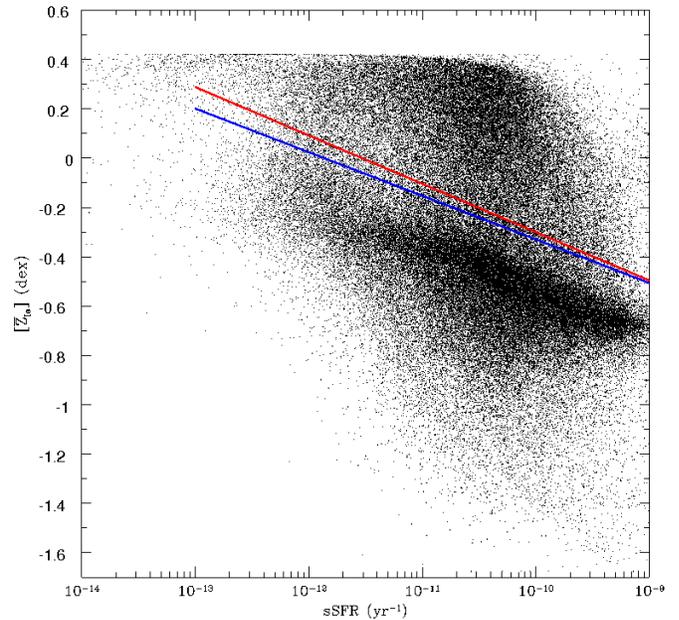}
\caption{
Mean $\bar{Z}_\mathrm{Ia}$ as a function of the specific SFR, as derived from the VESPA catalogue
(runID=2) and the DTD of M10b. The solid lines are the results of linear fits to the points given
by $[\bar{Z}_\mathrm{Ia}] = -0.177\log\left(\mathrm{sSFR}\right) - 2.10$ (blue, early-type
galaxies) and $[\bar{Z}_\mathrm{Ia}] = -0.196\log\left(\mathrm{sSFR}\right) - 2.26$ (red, late-type
galaxies). 
}\label{figssfr}
\end{figure}

Figure~\ref{figssfr} shows $\bar{Z}_\mathrm{Ia}$ as a function of the sSFR (specific Star
Formation Rate). We first notice the sharp cut on top of the points distribution at
$\bar{Z}_\mathrm{Ia} = 0.4$, which derives from the maximum metallicity fed to the SSP models of
VESPA. If we disregard this limitation of the models, we can still fit a tendency line to the data
points, with the result that can be seen in the plot: SNIa exploding in high sSFR galaxies do have
smaller $\bar{Z}_\mathrm{Ia}$ than those that explode in passive galaxies. On the other hand, the
tendency lines for early and late-type galaxies do not differ significantly.
While the correlation is not strong, we
note that a relationship between sSFR and SNIa luminosity has been found in some observational
studies of SNIa \citep[e.g., to cite just one,][who found that SNIa in low sSFR galaxies appear
brighter on average than those in high sSFR galaxies, after applying stretch corrections]{sul10}.
As a note of caution, we remark that our galaxy sample was drawn from the whole unfiltered VESPA
catalogue whereas each observational study of SNIa properties are based on particular selection
function of hosts. Thus, a direct comparison with our results might not be meaningful.

\begin{figure}
\centering
   \includegraphics[width=9 cm]{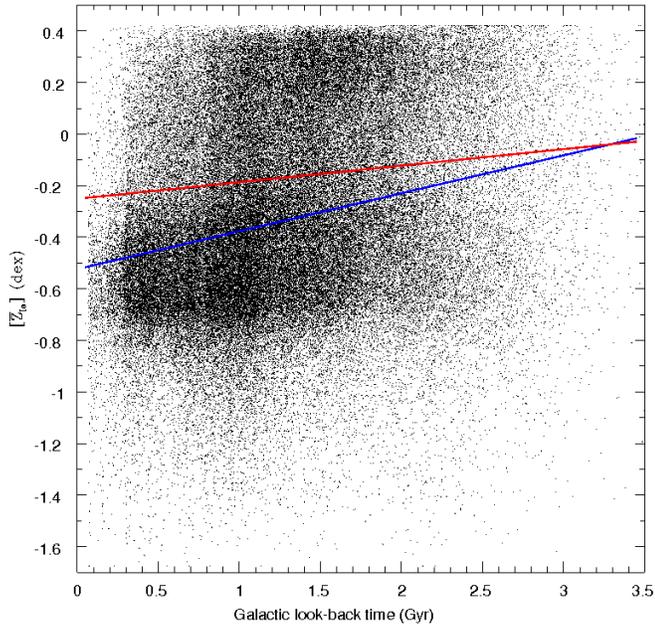}
\caption{
Mean $\bar{Z}_\mathrm{Ia}$ as a function of the look-back time of the host galaxy,
$t_\mathrm{back}$, derived from the VESPA catalogue (runID=2) and the DTD of M10b. The
solid lines are the results of linear fits to the points given by 
$[\bar{Z}_\mathrm{Ia}] = 0.147 t_\mathrm{back} - 0.524$ (blue, early-type galaxies) and 
$[\bar{Z}_\mathrm{Ia}] = 0.064 t_\mathrm{back} - 0.251$ (red, late-type galaxies).
}\label{figtback}
\end{figure}

Figure~\ref{figtback} shows the distribution of $\bar{Z}_\mathrm{Ia}$ as a function of the
look-back time of the host galaxy (that is: $t_\mathrm{back} = d/c$, where $d$ is the distance to
the galaxy and $c$ is the speed of light), that is itself a measure of the age of the universe at
the local host time, and of the redshift. Besides the accumulation of points around $Z=0.04$, the
data show a tendency for smaller $\bar{Z}_\mathrm{Ia}$ at lower redshift. The trend of
$Z_\mathrm{host}$ with look-back time (not shown in the Fig.) is opposite to that of
$\bar{Z}_\mathrm{Ia}$, i.e. older hosts are less metal-rich as expected from popular
galaxy evolution models. 
The sign of the slope of $\bar{Z}_\mathrm{Ia}$ vs. $t_\mathrm{back}$ is difficult to explain, as it
implies that SNIa are on average {\it less} luminous at high redshifts, contrary to what is
derived from observational samples \citep[e.g.][their Fig.~8]{how09}. 
Anyway, we cannot derive any strong conclusion given the limitations of the
present analysis, as the trend we detect might be a result of observational biases of the
catalogued data. 

We have checked that the above trends are qualitatively independent of the assumptions concerning
the DTD and the specific runs of VESPA that can be selected (galactic samples, SSP models, IMF
prescriptions, and dust models).
We have also tested the scatter of the parameters of the tendency lines in
Figs.~\ref{figssfr} and \ref{figtback} as a result of the uncertainties in the galactic properties
stored in the catalogue (errors in the metallicity and star formation rates in each galactic bin).
To this end, we have computed again the tendency lines for 200 random realizations of the galactic
sets (early-types and late-types) by adding random noise according to the catalogued metallicity
and SFR errors in every time bin of each galaxy of our VESPA sample. For each random
realization, we have computed the
tendency line and thereafter we have analyzed the distribution of slopes, their mean and standard
deviation. The standard deviation of the slope of the tendency lines of sSFR and $t_\mathrm{back}$
with respect to $\bar{Z}_\mathrm{Ia}$ is typically less than 0.002. 

\section{Conclusions}\label{toconclude}

We have calculated the chemical evolution of homogeneous galaxy models together with the evolution
of the supernova rates in order to evaluate the Metallicity Distribution Function of SNIa,
$\mathrm{MDF}(\Delta Z)$, i.e. the probability that the logarithm of the metallicity of a SNIa
exploding now differs in less than $\Delta Z$ from that of its host. We have analysed several model
galaxies aimed to represent from active to passive galaxies, including dwarf galaxies prone to
experience supernova driven outflows. 

Our results show a remarkable degree of coincidence (in an
statistical sense) between the mean $Z_\mathrm{Ia}$ and $Z_\mathrm{host}$. The dispersion of
$Z_\mathrm{Ia}$ in active galaxy models is quite small, meaning that $Z_\mathrm{host}$ is a quite
good estimator of the supernova metallicity, while passive galaxies present a larger dispersion. We
have devised a procedure to correct the difference in
metallicity between SNIa and their hosts (Sect.~\ref{sectstat}), based on a linear fit given in
Fig.~\ref{figZiaP}. These results
are insensitive to the choice of IMF, stellar lifetime and
stellar yields, within the ranges we have explored. In contrast, the Delay Time Distribution of
SNIa remains one of the main ingredients influencing the difference of metallicity between SNIa and
their hosts.

We have discussed the use of different metallicity indicators (Fe vs. O, gas-phase metallicity vs.
stellar metallicity). Using O (assuming its evolution is representative of that of the CNO group)
as a metallicity measure for late-type galaxies does not change appreciable the metallicity
correction with respect to using Fe (assuming its evolution is representative of that of all
metals)\footnote{The precise meaning of this statement is that the mass fraction of CNO elements in
SNIa progenitors is represented by their mass fraction in the host galaxies, with similar accuracy
as the whole metal mass fraction in the progenitors is represented by the whole metallicity of the
hosts.}. It is remarkable that the metallicity correction for early-type galaxies
($\tau_\mathrm{sfr}<1$~Gyr) is quite different for O than for Fe. On the other hand, the
metallicities of SNIa are much better represented by those of the gas-phase than by the mean
stellar metallicities. Thus, when possible, it is advisable to measure the host metallicity through
that of its gas-phase.

Finally, the results of the application of our formalism to a galactic catalogue (VESPA) suggest
that SNIa come, in average, from
smaller metallicity progenitors both at low redshifts and in galaxies with large star-formation
activity. The paucity of metallicities used in the original stellar-population synthesis models
used in the construction of the galactic catalogue does not allow us to build a significant MDF of
SNIa based on the galactic histories stored in the database. However, 
in spite of the large uncertainty in the metallicity derived from the catalogue, the gross trends of $\bar{Z}_\mathrm{Ia}$ vs. $Z_\mathrm{host}$ obtained from VESPA for different galaxy types are roughly consistent with our theoretical estimates.
The derivation of improved,
observationally based, MDF will be possible in the future if the SSP models include further
refinements in their grid of metallicities.

One of the main drawbacks of our approach is the use of homogeneous one-zone galactic chemical
evolution models. Actual galaxies are heterogeneous, and radial gradients of chemical composition
are routinely measured within them. A possible improvement of our model would be to divide the
galaxy in several non-interacting shells as in, e.g., \citet{mar98}, then computing the galactic
MDF accounting for the probability that a SNIa originates in each shell. Then, however, a
complication would arise when comparing to observations of far away galaxies for which there is no
possibility of measuring the metallicity gradient: what is the meaning of {\it the} galaxy
metallicity in a context where each independent shell has its own chemical history?
What is clear is that the derived MDF of SNIa would be characterized by a larger dispersion than
the MDF we have obtained. In some sense, the true MDF of SNIa reflects a double dependence: in
space, i.e. the
position of the SNIa progenitor within its host galaxy, and in time, i.e. the moment in the
galactic history when the progenitor was born. It is just this last dependence what we have
studied in the present work.

\section*{Acknowledgments}

We thank the referee for his/her comments that have helped to improve substantially the present
study.
This work has been partially supported by a MEC grant, by
the European Union FEDER funds, and by the Generalitat de Catalunya. CB
thanks Benoziyo Center for Astrophysics for support. 
    Funding for the SDSS and SDSS-II has been provided by the Alfred P. Sloan Foundation, the
Participating Institutions, the National Science Foundation, the U.S. Department of Energy, the
National Aeronautics and Space Administration, the Japanese Monbukagakusho, the Max Planck Society,
and the Higher Education Funding Council for England. The SDSS Web Site is http://www.sdss.org/.
    The SDSS is managed by the Astrophysical Research Consortium for the Participating
Institutions. The Participating Institutions are the American Museum of Natural History,
Astrophysical Institute Potsdam, University of Basel, University of Cambridge, Case Western Reserve
University, University of Chicago, Drexel University, Fermilab, the Institute for Advanced Study,
the Japan Participation Group, Johns Hopkins University, the Joint Institute for Nuclear
Astrophysics, the Kavli Institute for Particle Astrophysics and Cosmology, the Korean Scientist
Group, the Chinese Academy of Sciences (LAMOST), Los Alamos National Laboratory, the
Max-Planck-Institute for Astronomy (MPIA), the Max-Planck-Institute for Astrophysics (MPA), New
Mexico State University, Ohio State University, University of Pittsburgh, University of Portsmouth,
Princeton University, the United States Naval Observatory, and the University of Washington.

\end{document}